%% file: iclr2027_conference.tex
\documentclass{article} 
\usepackage{iclr2027_conference,times}

\input{math_commands.tex}

\usepackage{hyperref}

\usepackage[utf8]{inputenc} 
\usepackage[T1]{fontenc}    
\usepackage{booktabs}       
\usepackage{amsfonts}       
\usepackage{nicefrac}       
\usepackage{microtype}      
\usepackage{xcolor}         
\usepackage{graphicx,booktabs,tabularx,multirow,amsmath,amssymb,xcolor}
\usepackage[capitalize,noabbrev]{cleveref}
\usepackage{wrapfig}
\usepackage{subcaption}
\providecommand{\cmark}{\ensuremath{\checkmark}}
\providecommand{\xmark}{\ensuremath{\times}}
\providecommand{\pmark}{\ensuremath{\triangle}}

\title{The Evaluation Substrate Matters: \\
{\fontsize{17pt}{18pt}\selectfont
Benchmarking Multi-Agent Serving Substrates}}

\author{
Haoyuan Zheng\textsuperscript{1}\thanks{Equal contribution.}\, , 
Dongting Li\textsuperscript{1}\footnotemark[1]\, , 
Hancheng Ye\textsuperscript{1},
Qinsi Wang\textsuperscript{1},
Mingyuan Ma\textsuperscript{1}, \\[0.5ex]
\textbf{
Yuncheng Yao\textsuperscript{1}, 
Yifei Wang\textsuperscript{1},
Baijun Zhou\textsuperscript{1},
Zishan Shao\textsuperscript{1},
Tingjun Chen\textsuperscript{1},
Hai Helen Li\textsuperscript{1},
} \\[0.5ex]
\textbf{
Danyang Zhuo\textsuperscript{2},
Yiran Chen\textsuperscript{1}
} \\[0.5ex]
\textsuperscript{1}
Department of Electrical and Computer Engineering\\
\textsuperscript{2}
Department of Computer Science\\[0.5ex]
Duke University\\
Durham, NC 27708, USA \\[0.5ex]
\texttt{\{haoyuan.zheng, hancheng.ye\}@duke.edu} 
}

\newcommand{\sys}{\textsc{MASSys}}

\newcommand{\zhy}[1]{}

\definecolor{revBlue}{HTML}{000000}
\definecolor{revOrange}{HTML}{000000}
\definecolor{revViolet}{HTML}{000000}

\newcommand{\revHMGh}[1]{{\color{revBlue}#1}}
\newcommand{\revtaDc}[1]{{\color{revOrange}#1}}
\newcommand{\revxXpx}[1]{{\color{revViolet}#1}}

\iclrfinalcopy 
\begin{document}

\maketitle
\lhead{Preprint}

\input{Sections/0_abstract}
\input{Sections/1_introduction}
\input{Sections/2_related_work}
\input{Sections/3_MASSys_benchmark}
\input{Sections/4_Experiments}
\input{Sections/5_conclusion}

{
    \small
    \bibliographystyle{plainnat}
    \bibliography{references}
}

\input{Sections/6_appendix}

\end{document}

%% file: math_commands.tex
\usepackage{amsmath,amsfonts,bm}

\def\eqref#1{equation~\ref{#1}}

\def\1{\bm{1}}

\DeclareMathAlphabet{\mathsfit}{\encodingdefault}{\sfdefault}{m}{sl}
\SetMathAlphabet{\mathsfit}{bold}{\encodingdefault}{\sfdefault}{bx}{n}



%% file: Sections/0_abstract.tex
\begin{abstract}

Multi-agent systems (MAS) usually coordinate heterogeneous models and tools to solve complex tasks, creating a growing need for reliable evaluation. However, a multi-agent workflow that is fastest in one serving environment can lose its advantage in another, even when its models, prompts, and task remain fixed. This makes a benchmark result conditional on how the workflow is executed. We introduce \sys{}, a benchmark for measuring this dependence in multi-agent large language model (LLM) systems. \sys{} executes the same candidate--task pairs across serving substrates and load regimes, using five workload families spanning software repair, research synthesis, financial analysis, interactive control, and diagram generation. It records task quality alongside delivery, latency, cost, resource pressure, and failure evidence. In a controlled motivation diagnostic, the Star workflow falls from first to fourth in mean-latency rank under KV-cache pressure, while Chain rises from fourth to first. The substrate sweeps further show that increasing load creates long runtime tails and timeout pressure. In several task families, quality among evaluated outputs changes less than effective quality over submitted jobs: the system delivers fewer usable results even when the surviving outputs remain competitive. Tool and runtime traces reveal that these outcomes involve both model serving and non-model execution. Together, the results motivate evaluating MAS designs over explicit operating conditions and reporting which tasks are delivered, how well they are solved, and where execution breaks down.
\footnote{Code: \url{https://github.com/MAS-Sys-Bench/MAS-Sys} \\
Project Page: \url{https://mas-sys-bench.github.io/MAS-Sys-Page/} \\
Trace Dataset: \url{https://huggingface.co/datasets/MAS-Sys/MAS-Sys-Traces}}
\end{abstract}

%% file: Sections/1_introduction.tex
\section{Introduction}
\label{sec:introduction}

Large language models (LLMs) are increasingly deployed as components inside agentic systems that plan, call tools, exchange intermediate results, and coordinate multiple specialized roles. Modern multi-agent systems (MAS) make this trend explicit: a user request may be decomposed by a planner, dispatched to code, search, finance, visualization, or critique agents, and then merged by a verifier or manager. Frameworks such as AutoGen and CAMEL popularized conversational and role-based multi-agent programming, while graph- and state-oriented frameworks make it practical to build long-running, tool-using, multi-actor workflows \citep{wu2024autogen,li2023camel}. In parallel, agent benchmarks such as AgentBench, GAIA, SWE-Bench, and MultiAgentBench have shifted evaluation beyond static question answering toward interactive environments, tool use, software engineering, and collaboration or competition among agents \citep{liu2023agentbench,mialon2024gaia,jimenez2024swebench,zhu2025multiagentbench}.

This progress changes what a benchmark score means. For a multi-agent application, the score is no longer only a property of the underlying LLM: it also depends on topology, role prompts, model allocation, tool access, memory, error handling, and orchestration. Recent system-level evaluation work makes this point explicit. MASEval~\citep{emde2026maseval}, for example, argues that the complete agentic system should be the unit of analysis, and shows that framework and implementation choices can materially affect multi-agent benchmark results. This is an important step beyond model-centric leaderboards. However, it leaves a lower layer largely implicit: the \emph{serving substrate} on which the multi-agent system is executed. An agent can fail a coding benchmark before writing any solution code. On the Terminal-Bench task \texttt{bn-fit-modify}, Anthropic reports that installing a data-science stack can exhaust a tightly constrained container's memory. Across resource configurations, its Terminal-Bench 2.0 score differed by six percentage points with the same model, harness, and tasks~\citep{segato2026infrastructurenoise}. The execution environment can therefore determine whether a chosen solution strategy can even proceed.

The same issue appears across successive turns. Manus reports an average input-to-output token ratio of approximately 100:1, making reuse of accumulated context consequential for latency and cost~\citep{ji2025contextengineering}. Continuum studies the related serving problem: an agent pauses to run a tool, and retaining its KV cache competes with admitting other requests~\citep{li2025continuumstory}. These observations motivate treating resource allocation, queueing, cache management, and tool execution as part of the conditions under which an agent is evaluated.

\begin{figure*}[t]
    \centering
    \includegraphics[width=0.99\textwidth]{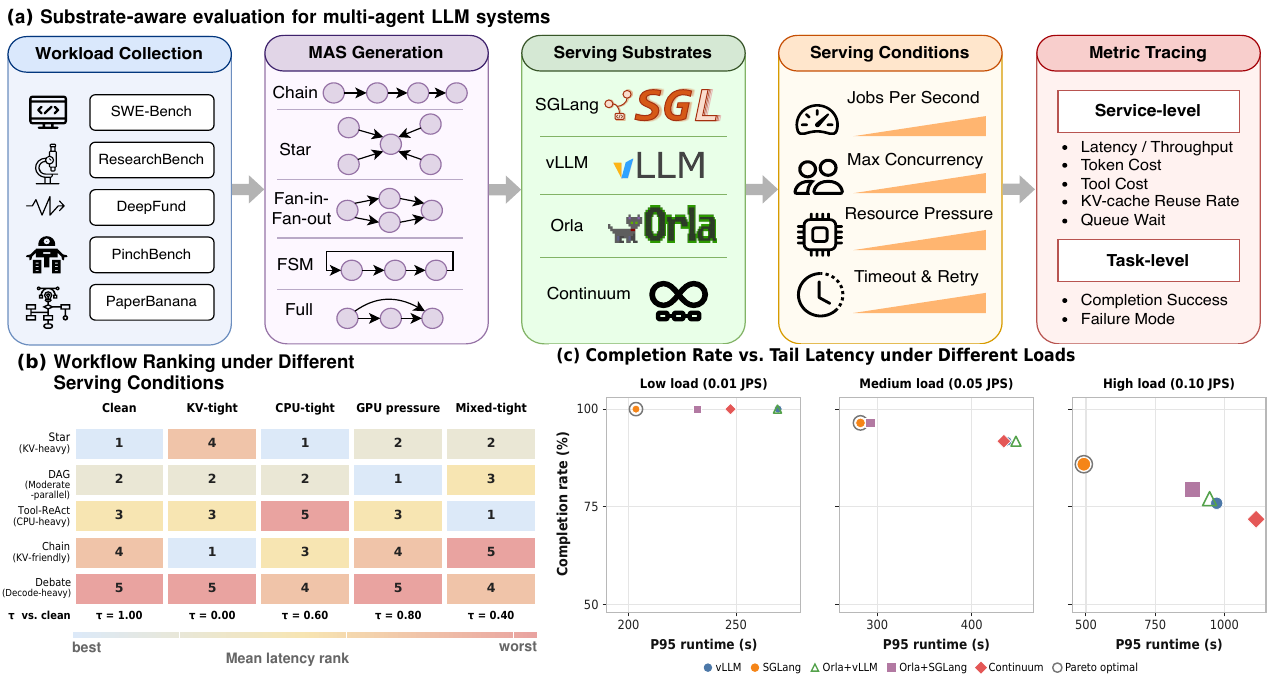}
    \caption{\textbf{Substrate-aware evaluation for multi-agent LLM systems.} (a) \sys{} fixes workloads and MAS candidates, then evaluates the same candidate-task pairs under different serving substrates and conditions. (b) A motivation diagnostic shows that workflow latency rankings change under controlled serving bottlenecks; Kendall's $\tau$ is reported only relative to the clean condition. \revHMGh{(c) Real 4$\times$H200 substrate-sweep samples show window completion rate versus run-level P95 runtime at 0.01, 0.05, and 0.10 JPS, with outer rings indicating Pareto-optimal points.}}
    \label{fig:intro-concept-protocol}
\end{figure*}

We use \emph{serving substrate} to refer to the runtime, backend, request load, queueing policy, resource budget, cache behavior, timeout/retry policy, and orchestration mechanism that mediate execution between a MAS design and the underlying models and tools. Existing agent and MAS benchmarks typically fix this layer and attribute score differences to the model, scaffold, or candidate system. Fig.~\ref{fig:intro-concept-protocol} illustrates why this is unsafe: the same workflow candidates change latency rankings under KV-cache, CPU, GPU, and mixed pressure, \revHMGh{while serving cells exhibit different window-completion and P95-runtime trade-offs across offered loads.}

We introduce \sys{}, a substrate-aware benchmark for multi-agent LLM systems. \sys{} draws tasks from SWE-Bench, ResearchBench, DeepFund, PinchBench\footnote{\href{https://github.com/pinchbench/skill}{https://github.com/pinchbench/skill}}, and PaperBanana, instantiates heterogeneous MAS candidates from a shared design space, and replays the same candidate-task pairs across interchangeable serving substrates and load conditions~\citep{jimenez2024swebench, yu2025researchtownsimulatorhumanresearch, li2025timetravelcheatinggoing, zhu2026paperbananaautomatingacademicillustration}. This paired design holds workload and MAS design fixed while making backend, runtime, load, resource pressure, and timeout behavior explicit evaluation variables. Table~\ref{tab:intro-positioning} displays the position of \sys{} among existing work. 

Our contributions are: (1) a substrate-aware benchmark protocol for MAS evaluation; (2) \sys{}, a mixed-domain benchmark with controlled candidates, interchangeable serving cells, and unified task/service traces; and (3) substrate-sweep evidence showing that backend/runtime choice, load, and workload composition change workflow rankings, tail latency, timeout behavior, tool pressure, and the observed reliability envelope.
\begin{table*}[t]
\centering
\setlength{\tabcolsep}{2.2pt}
\renewcommand{\arraystretch}{0.9}
\setlength{\abovecaptionskip}{2pt}
\setlength{\belowcaptionskip}{2pt}
\setlength{\aboverulesep}{0.25ex}
\setlength{\belowrulesep}{0.25ex}

\caption{\textbf{Positioning of \sys{}.} Comparison with representative benchmarks, frameworks, harnesses, and serving systems.}
\label{tab:intro-positioning}

\fontsize{8pt}{9pt}\selectfont
\renewcommand{\tabularxcolumn}[1]{m{#1}}
\begin{tabularx}{\textwidth}{@{}>{\raggedright\arraybackslash}X *{6}{c}@{}}
\toprule
\textbf{Line of work} &
\textbf{Domain} &
\textbf{Workload} &
\textbf{Substr.} &
\textbf{Stress} &
\textbf{Dual} &
\textbf{Trace} \\
\midrule

\textbf{General evaluation suites}:
HELM, OpenCompass, Inspect-AI, lm-eval, LightEval~\citep{liang2022helm,opencompass2023,inspectai2024,biderman2024lmeval,lighteval2024}
& \cmark & \xmark & \xmark & \xmark & \pmark & \pmark \\
\addlinespace[2pt]

\textbf{Agent and tool-use benchmarks}:
AgentBench, GAIA, SWE-Bench, WebArena, OSWorld, AppWorld, ToolBench, BFCL, $\tau^2$-Bench~\citep{liu2023agentbench,mialon2024gaia,jimenez2024swebench,zhou2023webarena,xie2024osworld,trivedi2024appworld,qin2023toolllm,patil2025bfcl,barres2025tau2bench}
& \cmark & \pmark & \xmark & \xmark & \pmark & \pmark \\
\addlinespace[2pt]

\textbf{Multi-agent benchmarks}:
MultiAgentBench, MARBLE, ChatEval, ConVerse~\citep{zhu2025multiagentbench,chan2023chateval,gomaa2025converse}
& \cmark & \cmark & \xmark & \xmark & \pmark & \pmark \\
\addlinespace[2pt]

\textbf{Multi-agent frameworks}:
AutoGen, CAMEL, LangGraph, CrewAI, LlamaIndex, AgentVerse, AgentScope, MetaGPT, ChatDev~\citep{wu2024autogen,li2023camel,langgraph2024,crewai2024,llamaindex2022,chen2023agentverse,gao2024agentscope,hong2023metagpt,qian2024chatdev}
& \pmark & \cmark & \xmark & \xmark & \xmark & \xmark \\
\addlinespace[2pt]

\textbf{Agent harnesses}:
AutoGenBench, HAL, AnyAgent, AgentGym, MCPEval~\citep{fourney2024autogenbench,kapoor2026hal,mozilla2025anyagent,xi2024agentgym,lastmile2025mcpeval}
& \pmark & \pmark & \pmark & \pmark & \pmark & \pmark \\
\addlinespace[2pt]

\textbf{LLM-app evaluation}:
DeepEval, Phoenix, TruLens, MLflow GenAI, LangSmith, RAGAs, Promptfoo, Langfuse, Weave~\citep{deepeval2026,arize2022phoenix,trueera2020trulens,zaharia2018mlflow,langsmith2024,es2024ragas,promptfoo2024,langfuse2024,weave2024}
& \pmark & \xmark & \xmark & \xmark & \pmark & \pmark \\
\addlinespace[2pt]

\textbf{System-level MAS evaluation}:
MASEval, MAFBench, AgentTaxo~\citep{emde2026maseval,orogat2026mafbench,wang2025agenttaxo}
& \cmark & \cmark & \pmark & \pmark & \pmark & \pmark \\
\addlinespace[2pt]

\textbf{Generic LLM serving}:
vLLM, SGLang, Orca, DistServe, Sarathi-Serve, TensorRT-LLM, TGI, LightLLM~\citep{kwon2023vllm,zheng2024sglang,yu2022orca,zhong2024distserve,agrawal2024sarathi,nvidia2023tensorrtllm,huggingface2023tgi,lightllm2023}
& \xmark & \xmark & \cmark & \cmark & \pmark & \pmark \\
\addlinespace[2pt]

\textbf{Agentic/application serving}:
Parrot, Ayo, Orla, Continuum, Helium, Nalar, Autellix, SDAS~\citep{lin2024parrot,tan2025ayo,shahout2026orla,li2025continuumstory,wadlom2026helium,laju2026nalar,luo2026agentix,agarwal2026sdas}
& \pmark & \pmark & \cmark & \cmark & \pmark & \pmark \\

\midrule
\textbf{\sys{}} & \cmark & \cmark & \cmark & \cmark & \cmark & \cmark \\
\bottomrule
\end{tabularx}

\begin{minipage}{0.99\textwidth}
\fontsize{7pt}{8pt}\selectfont
\textit{Legend.} \cmark{} = primary objective; \pmark{} = partially covered or indirectly available; \xmark{} = not a primary evaluation dimension.
\par\smallskip
\textit{Dimensions.} Domain = multi-domain coverage; Workload = variation of topology, agent count, prompts, models, and tools; Substr. = backend/runtime as explicit variable; Stress = load, queueing, timeout, failure, or resource pressure; Dual = task- and service-level metrics; Trace = failure-mode tracing under paired changes.
\end{minipage}
\end{table*}

%% file: Sections/2_related_work.tex
\section{Related Work}
\label{sec:related}

\begin{figure}[t]
    \centering
    \includegraphics[width=0.93\linewidth,trim=0 6bp 0 5bp,clip]{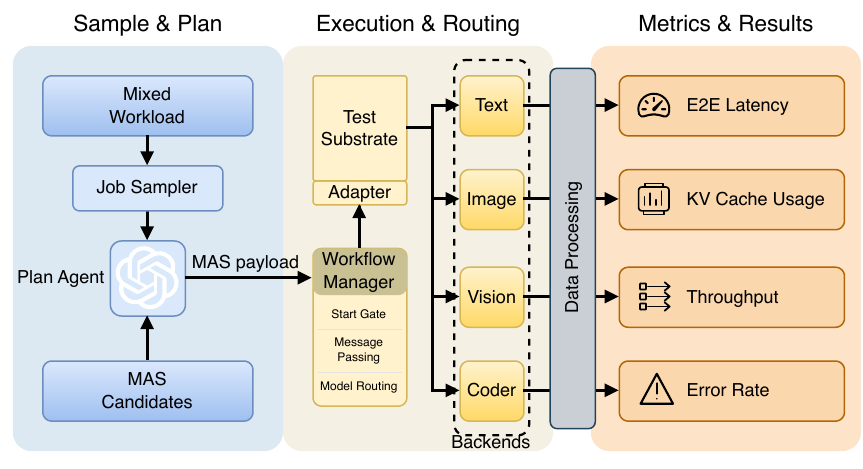}
    \captionsetup{skip=4pt}
    \caption{\textbf{\sys{} evaluation architecture.} \sys{} samples mixed workloads and MAS candidates, converts them into a structured MAS payload through a plan agent, and executes the payload inside a substrate-aware routing layer. The workflow manager coordinates start gates, message passing, and model routing across text, image, vision, and coding backends. The resulting traces are processed into service-level metrics such as end-to-end latency, KV-cache usage, throughput, and error rate, enabling controlled comparison of MAS behavior across serving substrates.}
    \label{fig:massys-architecture}
\end{figure}


\paragraph{Agent and MAS benchmarks.}
Agent benchmarks have moved LLM evaluation from static question answering to executable tasks with tools, environments, and long-horizon interaction. AgentBench and GAIA evaluate general agent abilities; SWE-Bench, WebArena, OSWorld, AppWorld, ToolBench, BFCL, and $\tau^2$-Bench stress software repair, web or desktop control, app-level actions, tool/API use, function calling, and conversational control~\citep{liu2023agentbench,mialon2024gaia,jimenez2024swebench,zhou2023webarena,xie2024osworld,trivedi2024appworld,qin2023toolllm,patil2025bfcl,barres2025tau2bench}. Multi-agent benchmarks add collaboration, competition, debate, safety, and organization structure through datasets such as MultiAgentBench, ChatEval, ConVerse, and AgentTaxo~\citep{zhu2025multiagentbench,chan2023chateval,gomaa2025converse,wang2025agenttaxo}. These works provide the task side of our benchmark, but they usually hold the execution substrate fixed. \sys{} keeps the MAS candidate and task distribution fixed while varying the backend, runtime, load, queueing, and resource conditions that execute the candidate.


\paragraph{Evaluation infrastructure and system-level MAS evaluation.}
General evaluation suites such as HELM, OpenCompass, Inspect-AI, lm-eval, and LightEval improve reproducibility and comparability for language-model evaluation~\citep{liang2022helm,opencompass2023,inspectai2024,biderman2024lmeval,lighteval2024}. LLM-application tools such as DeepEval, Phoenix, TruLens, MLflow GenAI, LangSmith, RAGAs, Promptfoo, Langfuse, and Weave add tracing, regression tests, observability, and task-specific metrics~\citep{deepeval2026,arize2022phoenix,trueera2020trulens,zaharia2018mlflow,langsmith2024,es2024ragas,promptfoo2024,langfuse2024,weave2024}. Agent harnesses including AutoGenBench, HAL, AnyAgent, AgentGym, and MCPEval reduce the cost of repeated runs, sandboxing, framework integration, and tool or MCP evaluation~\citep{fourney2024autogenbench,kapoor2026hal,mozilla2025anyagent,xi2024agentgym,lastmile2025mcpeval}. The closest work is MASEval, which treats the complete multi-agent system (framework, topology, orchestration, tools, and model choices) as the evaluation unit; MAFBench similarly studies framework-level trade-offs~\citep{emde2026maseval,orogat2026mafbench}. \sys{} adopts this system-level view and moves one layer lower: after the MAS candidate is fixed, it measures whether substrate choices change task outcomes, service behavior, and failure evidence.

\paragraph{LLM serving and agentic/application serving.}
LLM serving systems make backend behavior a primary systems problem. vLLM, SGLang, Orca, DistServe, Sarathi-Serve, TensorRT-LLM, TGI, and LightLLM optimize memory management, structured-program execution, batching, prefill/decode scheduling, throughput, and latency~\citep{kwon2023vllm,zheng2024sglang,yu2022orca,zhong2024distserve,agrawal2024sarathi,nvidia2023tensorrtllm,huggingface2023tgi,lightllm2023}. Recent application and agentic-serving work, including Parrot, Ayo, Teola, Orla, Continuum, Helium, Nalar, Agentix/Autellix, and software-defined agentic serving, further targets workflow structure, KV-cache reuse, scheduling, and runtime control for multi-call or multi-turn applications~\citep{lin2024parrot,tan2025ayo,tan2024teola,shahout2026orla,li2025continuumstory,wadlom2026helium,laju2026nalar,luo2026agentix,agarwal2026sdas}. These systems optimize how workflows are served. \sys{} asks the evaluation question that follows: when a MAS benchmark is run on such substrates, which measured effects come from the candidate, the workload, or the substrate itself?

%% file: Sections/3_MASSys_benchmark.tex
\section{\sys{}: Substrate-aware MAS Evaluation Benchmark}
\label{sec:benchmark-description}



\sys{} is a benchmark and evaluation framework for studying how serving substrates affect multi-agent LLM systems. The benchmark is organized around three principles. First, the evaluated object is a complete MAS candidate, including its topology, number of agents, role prompts, model allocation, tools, and orchestration policy. Second, a serving substrate is a first-class experimental variable, covering backend/runtime choice, request load, queueing behavior, resource pressure, cache behavior, and scheduling policy. Third, every run records both task-level and service-level outcomes, so that a completed execution is not mistaken for a correct solution and a low-latency execution is not mistaken for a robust MAS design. Fig.~\ref{fig:massys-architecture} summarizes the evaluation pipeline that connects workload sampling, MAS candidate construction, substrate execution, and trace processing.

\sys{} evaluates complete MAS candidates under controlled serving substrates. Each candidate includes topology, agent count, role prompts, model assignment, tool access, memory, and orchestration policy. Each substrate cell specifies backend/runtime, offered load, queueing behavior, resource budget, cache policy, and timeout/retry policy. Every run records both task-level and service-level outcomes so that a completed execution is not mistaken for a correct answer and a low-latency execution is not mistaken for a robust MAS design.


\subsection{Workloads and Candidates}

\begin{wrapfigure}{R}{0.5\linewidth}
    \centering
    \includegraphics[width=\linewidth]{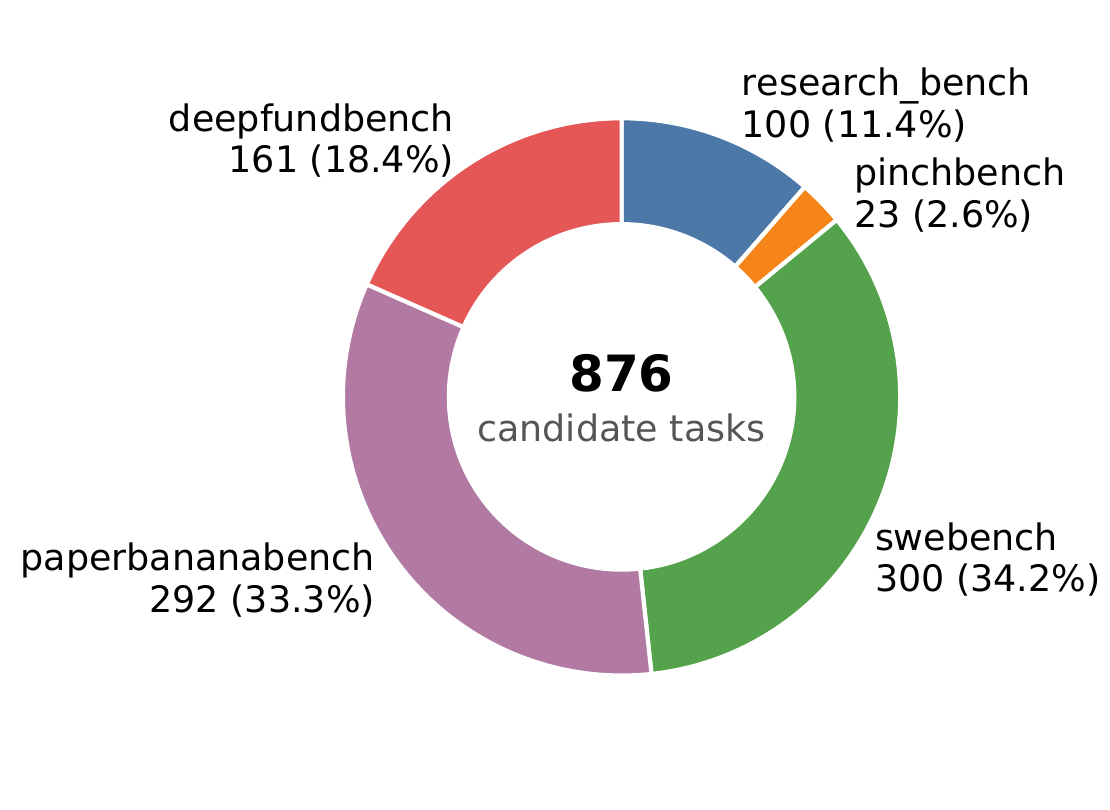}
    \caption{\textbf{Candidate task pool composition.} The pool contains 876 candidate tasks across five workload families.}
    \label{fig:candidate-pool}
\end{wrapfigure}

\begin{figure}[t]
    \centering
    \includegraphics[width=1\linewidth]{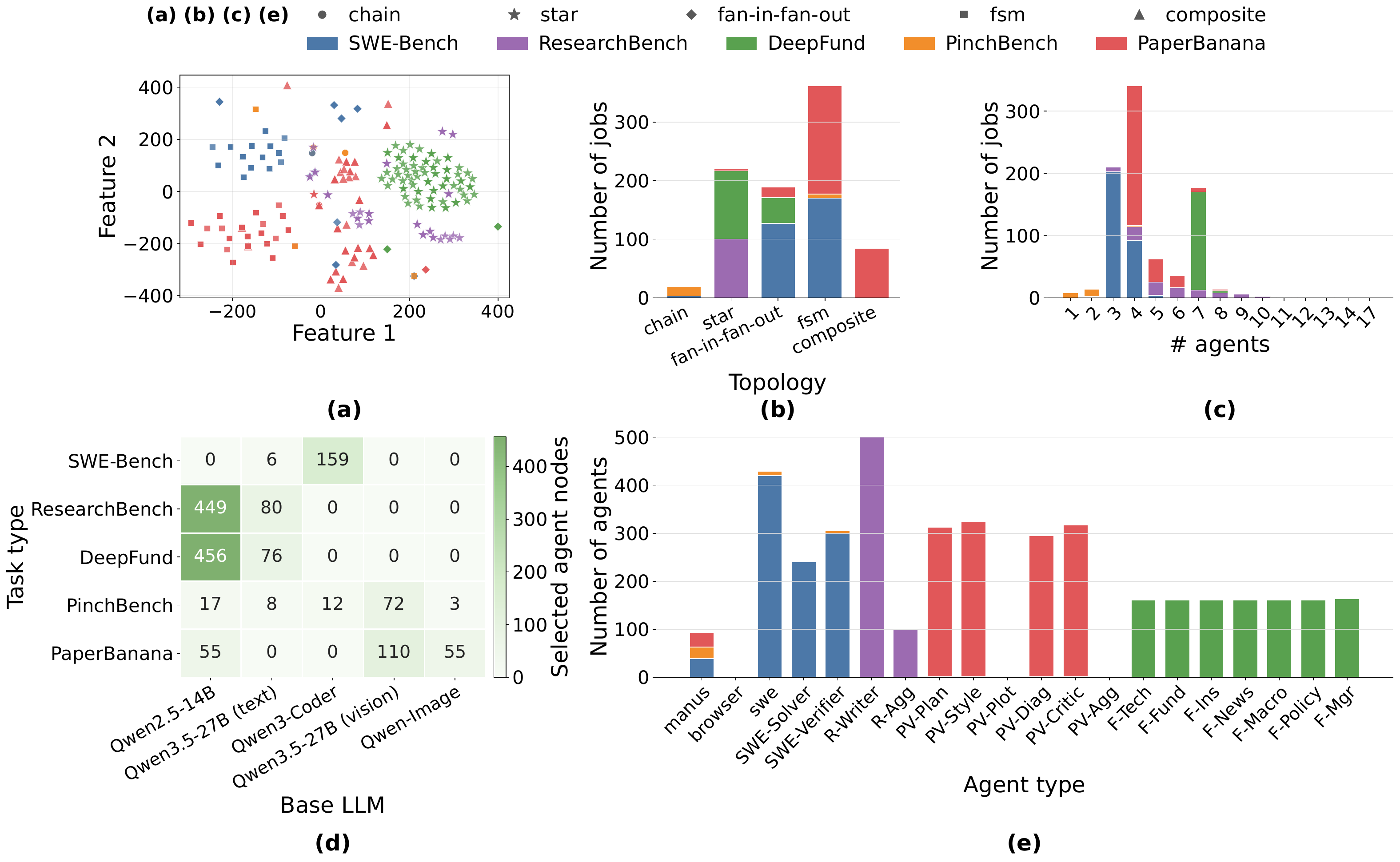}
    \caption{\textbf{MAS candidate topology, agent-role and base-LLM distribution.}}
    \label{fig:topology+agent+basellm}
\end{figure}

\sys{} uses five task families: SWE-Bench, ResearchBench, DeepFund, PinchBench, and PaperBanana. We evaluate them jointly as a mixed service workload instead of separate single-domain leaderboards. Each family has sampling probability 0.2; after selecting a family, the runner samples one task uniformly from its pool with replacement. The mixture covers software repair, research synthesis, financial analysis, interactive control, and diagram-oriented generation, producing jobs with different tool chains, fan-out/fan-in structures, verifier loops, and modality-specific routing. The candidate space covers chain, star, fan-in/fan-out, finite-state-machine, and composite workflows. Fig.~\ref{fig:candidate-pool} summarizes the candidate task pool, while Fig.~\ref{fig:topology+agent+basellm} summarizes the resulting MAS structure, aggregate agent-role distribution and generated topology diversity. Fig.~\ref{fig:topology+agent+basellm} further shows that model allocation is task-dependent across specialized backbones. Together, they show that \sys{} varies both benchmark source and candidate topology across a shared design space.

\subsection{Paired Substrate Sweeps}

\sys{} executes the same candidate--task pairs across vLLM-style serving, SGLang-style serving, Continuum-style serving, Orla over vLLM, and Orla over SGLang. The sweeps vary offered load and hardware regime while keeping the workload distribution fixed. This paired design makes the substrate an experimental variable: a change in completion, latency, timeout behavior, or failure evidence can be attributed to the serving condition without confounding from task-sampling shifts.

\subsection{Tracing and Metrics}

The trace layer records job status, end-to-end runtime, unfinished-at-window-close evidence, graph structure, tool calls and tool latencies, LLM-call counts, queue wait, token usage, concurrency, resource samples, and KV-cache behavior. Post-hoc evaluators compute task quality using benchmark-specific checks, including execution-based grading, structured validation, and judge-based scoring when needed. Table~\ref{tab:evaluation-metrics} summarizes the metric axes. \revHMGh{We distinguish runtime-threshold exceedance, execution success, deadline termination, and window completion. Fig.~\ref{fig:exp_completion_runtime_kv}(h) reports whether a job's end-to-end runtime exceeds a task-family-specific reference threshold: 50\,s for DeepFund, 130\,s for ResearchBench and PinchBench, 350\,s for SWE-Bench, and 360\,s for PaperBanana. These thresholds were selected by inspecting low-load runtime distributions and are applied uniformly across serving cells, loads, and hardware configurations. They are post-hoc reporting cutoffs and do not terminate execution. In contrast, every job is subject to an enforced 1{,}200\,s deadline. Window completion measures whether a job finishes within the 30-minute submission window; unfinished jobs may continue during the drain period. Thus, a job can exceed its runtime reference threshold and still achieve execution success.} This separation allows \sys{} to distinguish task quality, service feasibility, and failure evidence within the same run.

\begin{table}[t]
\small
\centering
\caption{\textbf{\sys{} Metrics.}}
\label{tab:evaluation-metrics}
    \begin{tabularx}{\linewidth}{>{\raggedright\arraybackslash}p{0.25\linewidth}X}
    \toprule
    \textbf{Service/System Metric} & \textbf{Definition} \\
    \midrule
    Completion & Fraction of submitted jobs that finish before the 30-minute submission window closes. \\
    Execution success & Fraction of submitted jobs whose worker returns a usable final artifact, independent of task correctness. \\
    \revHMGh{Runtime-threshold exceedance} & \revHMGh{Fraction of submitted jobs whose end-to-end runtime exceeds a post-hoc family reference cutoff (50\,s DeepFund; 130\,s ResearchBench/PinchBench; 350\,s SWE-Bench; 360\,s PaperBanana), without terminating execution.} \\
    \revHMGh{Deadline termination / execution failure} & \revHMGh{Fractions of submitted jobs that reach the enforced 1,200 s per-job deadline or fail to produce a usable final artifact.} \\
    Runtime and queueing & Mean/P95 end-to-end runtime and queue wait, computed over non-killed execution-success jobs. \\
    Cost and pressure & Token usage, tool latency, concurrency, and KV-cache pressure; KV-cache pressure is reported from backend samples as mean and P95. \\
    \midrule
    \textbf{Task-Quality Metric} & \\
    \midrule
    SWE-Bench & Docker harness resolved rate: resolved instances divided by evaluated instances. \\
    ResearchBench & Embedding cosine similarity between generated and reference research artifacts, averaged over evaluated jobs. \\
    DeepFund & Trading outcome score derived from realized or next-trading-day PnL labels: profit, flat, or loss. \\
    PinchBench & Combined score from automated graders and LLM-judge rubric scores, using the task's grading weights. \\
    PaperBanana & Overall visual score, computed as the mean of instruction faithfulness, conciseness, readability, and aesthetics judge scores, divided by 10. \\
    \bottomrule
    \end{tabularx}
\end{table}

%% file: Sections/4_Experiments.tex


\section{Experiments}
\label{sec:experiments}

We evaluate \sys{} on mixed MAS workloads under controlled substrate sweeps. The goal of this section is not to produce a single leaderboard of MAS candidates, but to characterize how backend, runtime, load, and orchestration conditions change the feasibility and trace behavior of the same class of multi-agent jobs. Unless otherwise stated, \emph{succeeded} denotes execution success: the job completed and returned a usable final artifact under the service condition.


\subsection{Evaluation Settings}
\label{sec:evaluation-settings}
We evaluate \sys{} with paired substrate sweeps over five serving cells: vLLM-style serving, SGLang-style serving, Continuum-style serving, Orla over vLLM, and Orla over SGLang. Table~\ref{tab:exp-setup-summary} summarizes the hardware-specific model allocation. \revtaDc{These two regimes represent capacity-matched deployment tiers instead of an isolated GPU-count ablation: both share Qwen-image-2512 for image generation and multimodal Qwen3.5-27B for vision and text on two GPUs, while 4$\times$H200 adds Qwen2.5-14B-Instruct and Qwen3-Coder-30B-A3B-Instruct on two additional GPUs. All substrate comparisons hold the model pool and arrival schedule fixed within each tier.} In the main sweeps (seed 42, 1{,}200\,s per-job deadline), MAS plans are generated by a \href{https://developers.openai.com/api/docs/models/gpt-5.4}{GPT-5.4} plan agent, and jobs arrive via a Poisson process over a 30-minute window at 0.005, 0.01, 0.02, 0.05, and 0.10 JPS on 2$\times$H200 (0.01--0.10 JPS on 4$\times$H200). We report completion, execution success, mean/P95 runtime, KV-cache pressure, \revHMGh{runtime-threshold exceedance, deadline termination}, run-completion elapsed time, and concurrency as separate service metrics.

\begin{table}[t]
\centering
\small
\caption{\textbf{Hardware and model sets for substrate sweeps.}}
\label{tab:exp-setup-summary}
    \begin{tabularx}{\linewidth}{l p{5cm} X X}
        \toprule
        \textbf{Hardware} & \textbf{Language Models} & \textbf{Vision Models} & \textbf{Image Models} \\
        \midrule
        2$\times$H200 &
        \href{https://huggingface.co/Qwen/Qwen3.5-27B}{Qwen3.5-27B} & Qwen3.5-27B & \href{https://huggingface.co/Qwen/Qwen-Image-2512}{Qwen-image-2512} \\
        4$\times$H200 &
        \href{https://huggingface.co/Qwen/Qwen2.5-14B-Instruct}{Qwen2.5-14B-Instruct}; Qwen3.5-27B; \href{https://huggingface.co/Qwen/Qwen3-Coder-30B-A3B-Instruct}{Qwen3-Coder-30B-A3B-Instruct} & Qwen3.5-27B & Qwen-image-2512\\
        \bottomrule
    \end{tabularx}
\end{table}

\subsection{Main Results}
\label{sec:main-results}

\begin{figure}[t]
    \centering
    \begin{subfigure}[t]{0.59\linewidth}
        \includegraphics[width=\linewidth]{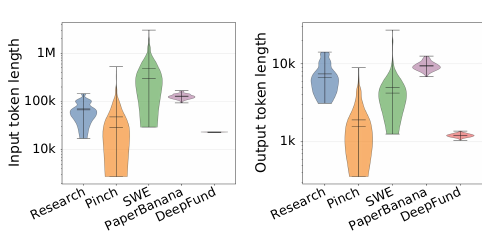}
        \caption{Token length.}
        \label{fig:exp_token_length}
    \end{subfigure}
    \hfill
    \begin{subfigure}[t]{0.39\linewidth}
        \includegraphics[width=\linewidth]{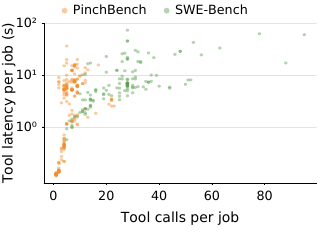}
        \caption{Tool latency vs. tool calls.}
        \label{fig:tool_cost}
    \end{subfigure}

    \caption{\textbf{Workload heterogeneity in token and tool demand. }(a) Input and output token length distributions vary substantially across task families, indicating different prompt/context sizes and generation scales. (b) Per-job tool latency generally increases with the number of tool calls but exhibits heavy tails, especially for SWE-Bench jobs with many edit, execution, and verification steps. These workload-level differences explain why mixed MAS evaluation stresses serving substrates differently from single-domain or single-request benchmarks.}
    \label{fig:load_sensitivity_serving_substrates}
\end{figure}

\begin{figure}[t]
    \centering
    \includegraphics[width=0.98\linewidth]{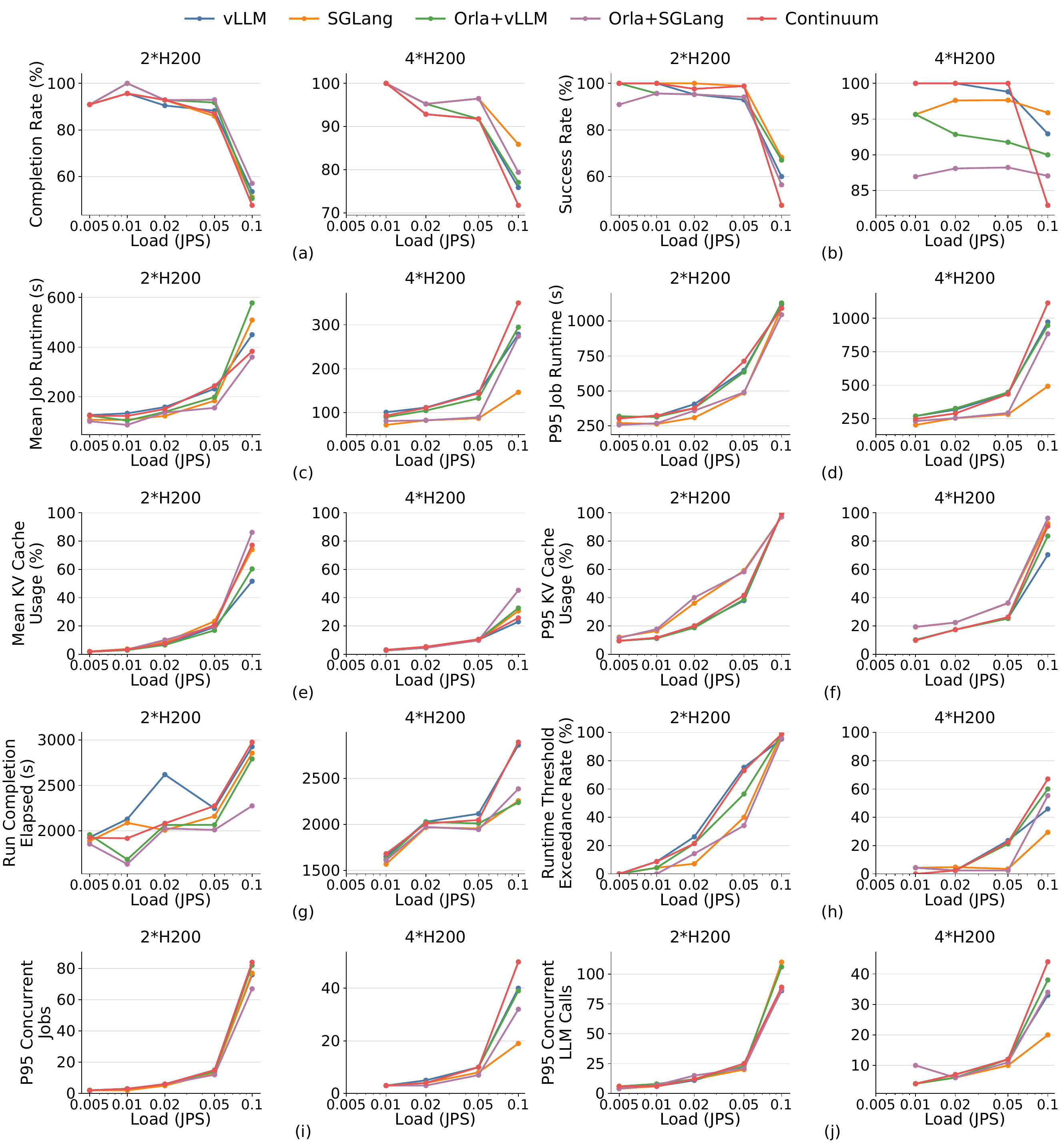}
    \caption{\textbf{Load-dependent service behavior across serving substrates.} Each pair of panels plots a service metric versus offered load for the 2$\times$H200 and 4$\times$H200 regimes, with color indicating the serving substrate. Completion and execution-success rates remain high at light load but degrade under stress, while mean and P95 runtime, KV-cache pressure, \revHMGh{runtime-threshold exceedance rate (post-hoc cutoffs of 50\,s DeepFund, 130\,s ResearchBench/PinchBench, 350\,s SWE-Bench, and 360\,s PaperBanana; enforced per-job deadline 1{,}200\,s)}, run-completion elapsed time, and P95 concurrency rise sharply as load increases. The figure shows that substrate comparisons are multi-dimensional: a serving cell may be fast at moderate load yet accumulate backlog, KV pressure, or failures under heavy mixed-MAS traffic.}
    \label{fig:exp_completion_runtime_kv}
\end{figure}

\paragraph{Result 1: Load turns MAS serving into a tail-latency problem.}
Increasing load affects the entire MAS execution graph beyond isolated model calls. Fig.~\ref{fig:exp_completion_runtime_kv} shows that, as offered load increases, completion and execution-success rates decline in several substrate cells, while mean job runtime, P95 runtime, \revHMGh{runtime-threshold exceedance rate}, run-completion elapsed time, KV-cache pressure, and P95 concurrency increase. The effect is visible in both 2$\times$H200 and 4$\times$H200 regimes. This pattern matters for MAS evaluation because multi-agent jobs contain multiple dependent calls, tool invocations, and aggregation steps. A delayed branch can block downstream agents or the final response, making tail behavior an integral part of the benchmark outcome.

\revHMGh{To assess statistical robustness across independent workload realizations (seeds 42--44), we added 50 runs covering all five serving cells and five loads on 2$\times$H200 and 12 runs covering native vLLM, SGLang, and Continuum at 0.05 and 0.10 jobs/s on 4$\times$H200 (Appendix~\ref{app:multiseed}). On 2$\times$H200, increasing load from 0.05 to 0.10 jobs/s reduces mean completion by 31.98--40.38 percentage points across cells (all five paired 95\% CIs exclude zero) and raises run-level P95 runtime in every cell (three of five CIs exclude zero). On 4$\times$H200, P95 runtime on the matched surviving-job intersection rises in all nine backend--seed comparisons (+152.0 to +302.9\,s on average), confirming consistent load-driven tail inflation across both deployment tiers.}

\paragraph{Result 2: backend/runtime choices change the service envelope.}
Fig.~\ref{fig:intro-concept-protocol}(c) and Fig.~\ref{fig:exp_completion_runtime_kv} show that substrate choice changes the completion-latency-reliability trade-off observed by the same mixed MAS workload. Fig.~\ref{fig:exp-task-quality} (Appendix~\ref{app:additional_figures}) shows that this service envelope also changes reported task quality. Effective quality is most sensitive in task families where serving failures remove many jobs from the evaluated set, with high-load drops especially visible for SWE-Bench, PinchBench, and PaperBanana in several 2$\times$H200 cells. By contrast, evaluated-only quality is often flatter for ResearchBench and parts of DeepFund, indicating that many completed-and-evaluated artifacts retain similar quality even as fewer jobs survive the service condition. The 4$\times$H200 regime mitigates some effective-quality loss but does not remove substrate dependence, especially for visual and interactive workloads. Some serving cells achieve lower runtime over portions of the load range, while others maintain stronger completion or execution-success behavior under the same offered load. The comparison is therefore not a single total ordering of backends. Instead, \sys{} exposes a service envelope: a substrate may be attractive for latency among completed jobs but less favorable in reliability, \revHMGh{runtime-threshold exceedance}, or backlog under heterogeneous mixed workloads. A deployment-relevant MAS benchmark should report these axes separately. \revtaDc{Substrate-dependent workflow ranking reversals also appear in the main benchmark. On PinchBench at 4$\times$H200 and 0.10 jobs/s, Chain achieves lower mean latency than FSM under vLLM, SGLang, and ORLA+SGLang, whereas FSM is faster under Continuum and ORLA+vLLM on the five-cell matched successful-job intersection (14 Chain and 7 FSM jobs; Table~\ref{tab:main_ranking_reversal}), and the cross-substrate Chain/FSM reversal recurs in seed 43 on the native 4$\times$H200 runs (Chain faster under SGLang, FSM faster under Continuum and vLLM; Appendix~\ref{app:main_workflow_ranking}).}

\revHMGh{We further audited all 3,255 submissions across the original 45 benchmark runs to distinguish execution failures from evaluation-infrastructure failures. Of the 492 jobs without valid evaluations, 490 resulted from execution failure or deadline termination, while only two encountered evaluation failures after successful execution. Overall, 2,763 of 2,765 execution-successful jobs received valid numerical scores. Thus, the missing-score component of the effective-quality gap stems almost entirely from execution and delivery failures instead of evaluator errors (Appendix~\ref{app:quality_audit}, Table~\ref{tab:evaluation_audit}).}

\revxXpx{The distinction between effective and evaluated-only quality also persists without LLM-based judging. In the original 2$\times$H200 runs, pooling the five serving cells, SWE-bench effective quality drops from 49.33\% to 11.72\% between 0.05 and 0.10 jobs/s, while evaluated-only quality changes from 64.91\% to 60.71\%. For automated-only PinchBench tasks, the corresponding changes are 93.81\% to 66.15\% and 93.81\% to 89.58\%. These execution- and artifact-grounded comparisons indicate that evaluated-only quality can substantially overstate useful quality delivered per submitted job (Appendix~\ref{app:quality_audit}, Table~\ref{tab:grounded_quality}).}

\paragraph{Result 3: workload composition changes the measured serving profile.}
Workload composition is itself a substrate-sensitive variable. Fig.~\ref{fig:load_sensitivity_serving_substrates} shows that task families differ in both input and output token length distributions, and that tool latency increases with the number of tool calls while exhibiting heavy tails. Fig.~\ref{fig:tool-results} (Appendix~\ref{app:additional_figures}) further decomposes the tool layer. Read/view, text-search, Python, test, and edit operations dominate call volume, but tool categories differ substantially in duration and success behavior. Browser and execution/testing paths are especially important because they combine non-trivial latency with more variable reliability. These observations explain why a mixed MAS workload can produce service behavior that is not predicted by agent count or request rate alone. Token distribution, tool mix, topology, and role allocation must be reported together with latency and completion metrics.



\paragraph{Result 4: trace decomposition identifies where substrate effects arise.}
Trace metrics reveal that the same final runtime can be caused by different mechanisms. Fig.~\ref{fig:decomposition} (Appendix~\ref{app:additional_figures}) decomposes mean critical-path latency across all five serving cells on 2$\times$H200 and 4$\times$H200 into backend prefill, backend decode, backend queue wait, backend other time, ORLA forwarding and backend time, tool execution, and agent/client overhead. At light and moderate load, critical-path time is dominated by backend decode together with tool execution and client overhead. At the highest load, backend queue wait and forwarding/client overhead become prominent. Combined with the tool-layer evidence in Fig.~\ref{fig:tool-results}, this decomposition shows why \sys{} records execution traces alongside scalar scores: slow or failed MAS executions can arise from backend decoding, queueing, tool calls, orchestration overhead, or their interaction along the MAS graph.
\subsection{Discussion}
\label{sec:discussion}


\paragraph{Fast substrates and robust substrates are not always the same.}
The results separate properties that are often conflated in benchmark reporting. A serving cell can have favorable runtime among completed jobs while showing weaker completion, \revHMGh{execution success, or backlog behavior, or higher runtime-threshold exceedance} under heavier load. Conversely, a substrate with higher latency may preserve a more stable service outcome in part of the operating range. This motivates reporting MAS benchmark results as service envelopes instead of a single backend or candidate score.

\paragraph{Backlog is an evaluation outcome.}
The fixed service-window view makes queue growth and unfinished work visible. Under heavy load, Fig.~\ref{fig:exp_completion_runtime_kv} shows that \revHMGh{runtime-threshold exceedance rate}, run-completion elapsed time, and concurrent-job pressure increase along with tail runtime. For interactive MAS applications, this is not a bookkeeping artifact. A delayed worker branch can block a verifier, a delayed verifier can block the final aggregation step, and a delayed final answer can turn an otherwise successful execution into an unusable service result. \sys{} therefore treats deadline behavior and unfinished-at-window-close evidence as first-class service metrics.

\paragraph{The tool layer is part of MAS serving.}
Fig.~\ref{fig:tool-results} shows that tool calls are neither uniform nor negligible. Some tools are frequent and lightweight; others are less frequent but dominate latency tails or reliability variation. This matters because MAS candidates differ in how they allocate work across search, reading, editing, execution, browser control, and verification. A model-only benchmark would miss this source of variation and may incorrectly attribute tool-layer effects to model quality or agent topology.

\paragraph{Failure attribution should be part of the benchmark report.}
Table~\ref{tab:evaluation-metrics} separates task outcome, service outcome, and failure evidence. This separation is necessary for substrate-aware evaluation. A failed MAS run may reflect task reasoning, agent coordination, tool behavior, environment constraints, queueing/timeout, resource pressure, backend/runtime behavior, or judge/format issues. \sys{} therefore records structured trace evidence and avoids collapsing all unsuccessful executions into a generic failure label. This design makes it possible to distinguish service-induced failures from model, tool, environment, or task-specific failures.

%% file: Sections/5_conclusion.tex
\section{Conclusion}
We presented \sys{}, a substrate-aware benchmark for multi-agent LLM systems. \sys{} executes controlled MAS candidates over mixed-domain workloads under paired serving substrates and records task outcomes, service outcomes, and trace-derived failure evidence. The reported sweeps show that workflow rankings change under resource bottlenecks, load creates heavy-tail runtime and timeout pressure, and serving cells expose different completion--latency--reliability envelopes. The workload and trace analyses further show that token distribution, tool mix, queueing, and orchestration overhead are essential parts of the benchmark outcome. These findings suggest that future MAS benchmarks should report substrate, load, hardware, workload structure, and trace evidence together with model and topology choices.

\section*{Ethics Statement}

This work collects no new human-subject data or private user interactions; its workloads derive from public benchmarks and data sources. Released traces are screened for credentials and deployment-specific identifiers, but may contain benchmark inputs and model-generated artifacts subject to upstream terms. DeepFund outputs are not financial advice, model-judge scores may inherit evaluator bias, and code/tool execution should use isolated environments with scoped credentials.

\section*{AI Use Statement}

In this work, we used generative AI tools to polish writing, retrieval and discovery, such as finding related work. We have not used generative AI tools for generating synthetic datasets, and proving mathematical claims are not applicable to this work. We have reviewed all AI-assisted work. We take responsibility for the final content of this work, including text, claims or artifacts produced with the aid of generative AI.

\section*{Reproducibility Statement}
Sections~\ref{sec:benchmark-description} and~\ref{sec:experiments} describe the benchmark protocol, workload construction, and experimental settings. Appendix~\ref{app:reproducibility} documents the execution, model and decoding, software, and evaluator configurations, together with replay-determinism scope. The accompanying code release provides the implementation, configurations, and execution artifacts used to derive the reported results.

%% file: Sections/6_appendix.tex
\newpage
\appendix

\section{Reproducibility}
\label{app:reproducibility}

\paragraph{Execution protocol.}
\revtaDc{Within each hardware--load--seed block, candidate MAS plans are generated once, and the identical task assignments, candidate configurations, and scheduled arrival offsets are replayed across all serving cells without replanning. Actual queueing, model responses, and tool execution remain independent across cells.}

\revHMGh{The original sweeps use seed 42; to quantify variability across independent workload realizations, we additionally executed seeds 43 and 44 across all five serving cells and five offered loads on 2$\times$H200 (50 new runs) and across the three native backends (vLLM, SGLang, and Continuum) at 0.05 and 0.10 jobs/s on 4$\times$H200 (12 new runs; Appendix~\ref{app:multiseed}).} The runner releases jobs at the frozen arrival offsets without a benchmark-level concurrency semaphore; native cells rely on the backend scheduler, while Orla uses per-backend concurrency and queue capacities of 1{,}000, chosen not to add a smaller front-end admission cap. The 30-minute window stops new arrivals and defines completion, but is not a global kill deadline: already submitted jobs drain until termination, subject to a 1{,}200\,s per-job deadline. Model-call transport errors are retried for at most six attempts with randomized exponential backoff between 1 and 60\,s.

\paragraph{Models and decoding.}
Each served model instance is deployed within one H200 (tensor parallelism one). We use BF16 weights without quantization and do not set an additional backend context-length cap, so the model/backend native context limit applies. Language and vision agents use temperature 0, a 4{,}096-token output cap, and disabled Qwen thinking mode; Qwen-image uses its image-generation defaults with temperature 1.0. The PlanAgent uses temperature 0 and an 8{,}192-token output cap. Backends start fresh for each run without an explicit warm-up, after which prefix caches accumulate within the run.

\paragraph{Software and artifacts.}
The released setup uses Python 3.11 and ORLA 1.2.15. The runtime environments use PyTorch/Transformers 2.11.0/5.6.2 for vLLM, 2.9.1/5.5.4 for SGLang, 2.10.0/4.57.6 for vLLM-Omni, and 2.8.0/5.6.2 for Continuum; the corresponding PyTorch builds target CUDA 13.0 or 12.8. vLLM is \texttt{0.19.2.dev0+gf150107.local} (commit \texttt{f150107}); the custom SGLang, vLLM-Omni, and Continuum source trees are included in the code release. Model identifiers are listed in Table~\ref{tab:exp-setup-summary}.

\revtaDc{The serving cells integrate existing frameworks with local compatibility modifications and do not independently reimplement their inference engines. ORLA retains its original scheduler with our runtime/client integration. For Continuum, we ported its scheduling, tool-duration estimation, and KV-retention mechanisms from the original vLLM 0.10.2-based release to a fork of the newer vLLM baseline used in our experiments so that Qwen3-family models share a matched underlying engine version across cells. Pinned package versions, launch configurations, and adapted backend source trees are available in the public code repository.}

\paragraph{Task-quality evaluators.}
ResearchBench uses \texttt{text-embedding-3-large} with 1{,}024-dimensional embeddings and cosine similarity. PinchBench uses the task's released automated grader, LLM rubric, and grading weights; its LLM judge is GPT-4o at temperature 0 and receives the task prompt, execution transcript, and task-specific rubric. PaperBanana uses GPT-4o at temperature 0 with the released PaperBanana prompts to score instruction faithfulness, conciseness, readability, and aesthetics on a 0--10 scale. SWE-Bench uses its per-instance Docker test harness. DeepFund maps the stored realized/next-trading-day outcome to profit/flat/loss and then to $1/0.5/0$. Evaluator prompts, rubrics, inputs, and structured outputs are retained in the code or trace releases.

Service aggregates are deterministically recomputable from the released traces. Repeating DeepFund against live data sources can change the market snapshot; the paper therefore uses the stored outcome labels. Exact commands, sanitized configurations, and artifact schemas are documented with the code and trace releases.

\section{Limitations}

\sys{} is intended to expose substrate effects rather than to exhaustively rank all MAS frameworks or serving systems. The current benchmark covers five workload families and a controlled set of MAS topologies, but it does not represent the full space of deployed agent applications, especially long-running enterprise workflows, human-in-the-loop tasks, and safety-critical interactive settings. The reported substrate sweeps also focus on a finite set of serving cells and hardware regimes; additional backends, schedulers, accelerators, and production admission-control policies may expose different trade-offs. Moreover, \sys{} separates execution success from task quality, but task-quality evaluation remains benchmark-specific and can depend on the fidelity of execution-based checks, structured validators, or judge models. Finally, failure attribution is trace-based rather than causal: the traces identify evidence for queueing, tool, resource, backend, or judge/format failures, but they may not fully isolate root causes when multiple failures interact along the same MAS graph. These limitations motivate extending \sys{} with broader workload sources, additional serving substrates, and stronger task-specific evaluators.

\section{Additional Task-Quality and Trace Results}
\label{app:additional_figures}

\begin{figure}[htbp]
    \centering
    \includegraphics[width=1\linewidth]{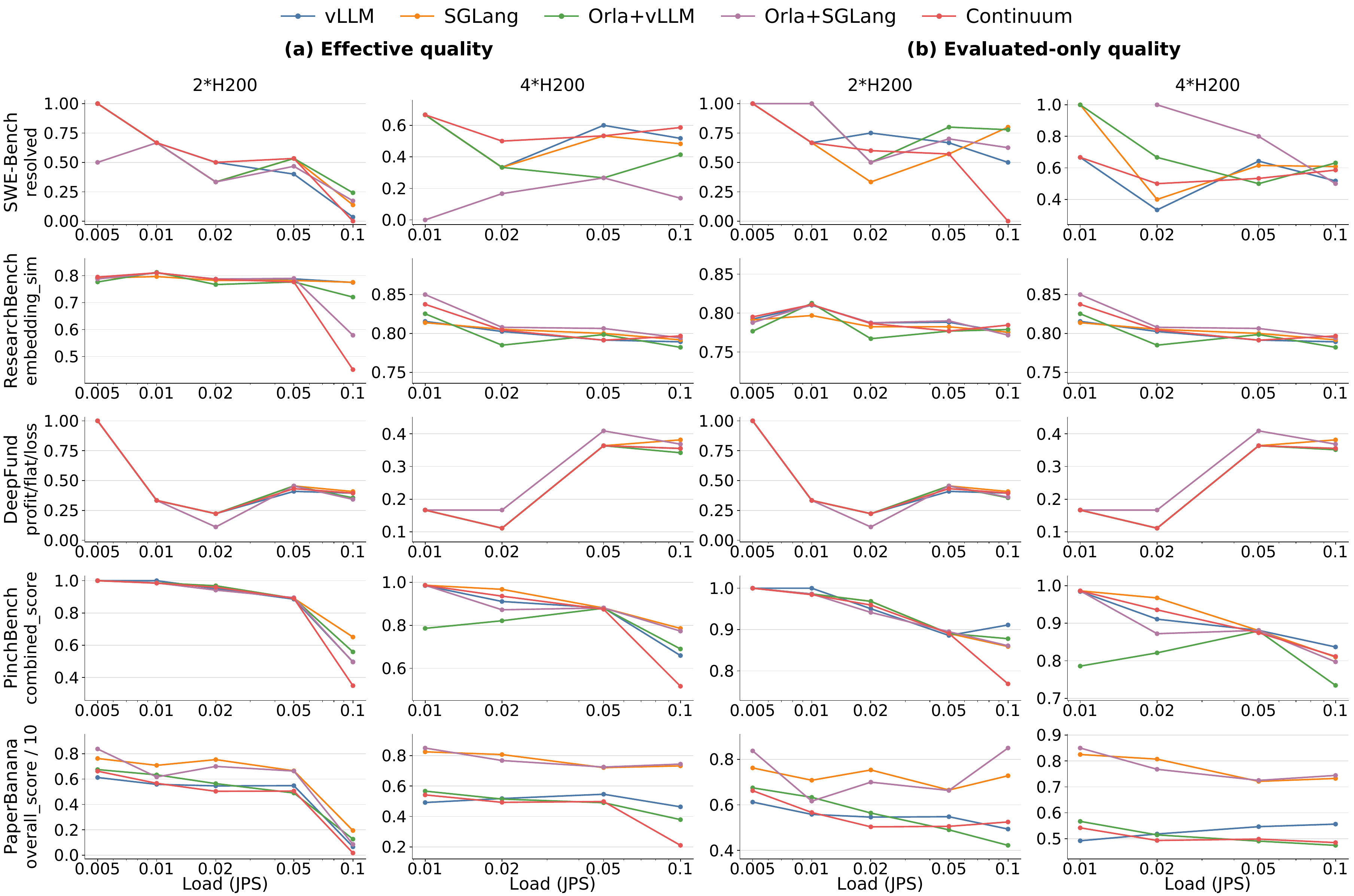}
    \caption{\textbf{Task quality across load and serving configurations.} The left half reports effective quality, where failed, deadline-terminated, or unevaluated jobs contribute zero, and the right half reports evaluated-only quality over jobs with valid post-hoc scores. Rows use each benchmark's normalized quality metric, columns compare 2$\times$H200 and 4$\times$H200 deployments, and curves show serving substrates across target JPS. \revHMGh{Auditing all 3{,}255 submissions confirms that 490 of 492 unscored jobs stem from execution or deadline failures and only two from post-success Docker pull errors (Appendix~\ref{app:quality_audit}, Table~\ref{tab:evaluation_audit}).}}
    \label{fig:exp-task-quality}
\end{figure}

\begin{figure}[htbp]
    \centering
    \includegraphics[width=1\linewidth]{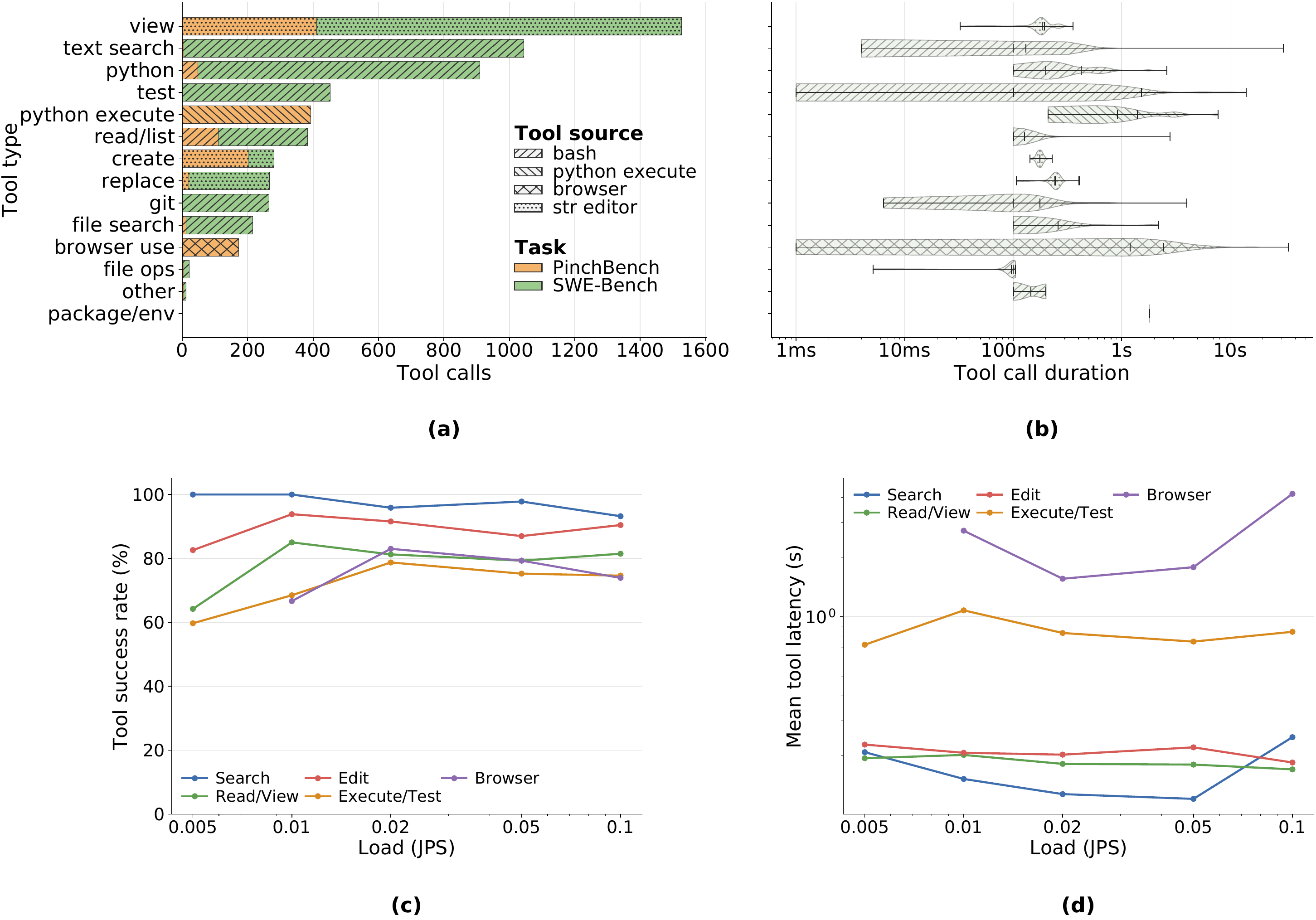}
    \caption{\textbf{Tool-layer profile of the mixed MAS workload.} (a) Tool calls are concentrated in read/view, text-search, Python, test, and edit operations, with SWE-Bench contributing most structured-editor and execution-heavy calls and PinchBench contributing browser/search activity. Hatching indicates the tool source. (b) Tool-call durations are highly type-dependent on a logarithmic scale: lightweight read/search/edit calls are short, whereas browser, test, git, and execution-related calls have heavier latency tails. (c) Tool success rates vary by category across load levels, with search remaining the most reliable and browser/edit/execution paths showing lower or more variable success. (d) Mean tool latency also depends strongly on category; browser and execute/test calls dominate tool-time cost and are the most sensitive to high offered load.}
    \label{fig:tool-results}
\end{figure}

\begin{figure}[htbp]
    \centering
    \includegraphics[width=1\linewidth]{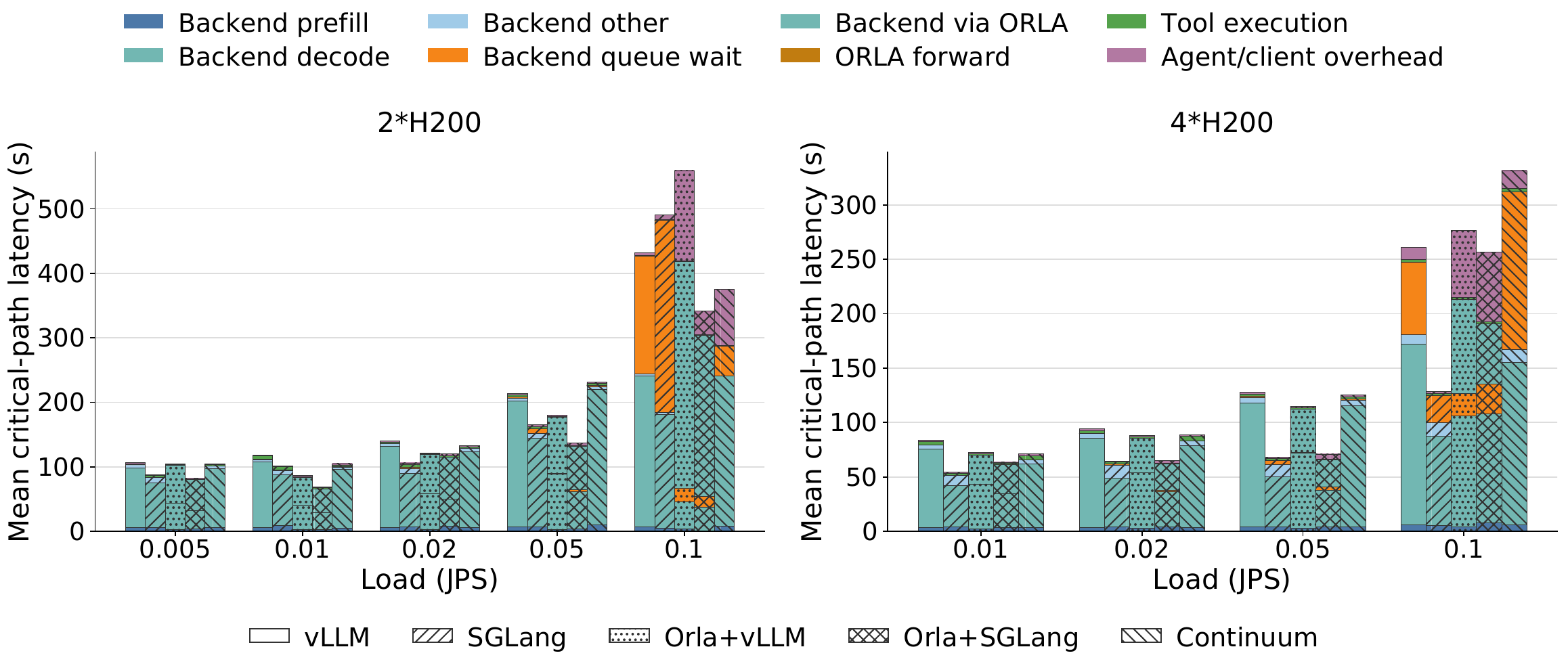}
    \caption{\textbf{Mean critical-path runtime decomposition across serving substrates on 2$\times$H200 and 4$\times$H200.}}
    \label{fig:decomposition}
\end{figure}

Figures~\ref{fig:exp-task-quality}--\ref{fig:decomposition} report the per-family task-quality curves, tool-layer latency and reliability profiles, and end-to-end runtime decomposition discussed in Section~\ref{sec:experiments}.

\section{\texorpdfstring{\revHMGh{Controlled Workflow-Ranking Experiment}}{Controlled Workflow-Ranking Experiment}}
\label{app:workflow_ranking}

\subsection{Experimental Setup}

\revHMGh{
Figure~1(b) presents a controlled workflow-ranking experiment that isolates the effects of serving resource constraints on multi-agent workflow latency. Unlike the main benchmark, this diagnostic experiment does not use PlanAgent-generated candidates or heterogeneous task workloads. Instead, five predefined workflows operate on a common synthetic telemetry context. The context is constructed once from deterministic numbered facts, with an approximate target length of 2{,}048 tokens, and reused across all serving conditions.
}

\revHMGh{
All workflows use Qwen2.5-14B-Instruct served by vLLM on a single NVIDIA H200 GPU, with tensor parallelism set to one, BF16 precision, a maximum model context of 8{,}192 tokens, and prefix caching disabled. Model requests use greedy decoding ($T=0$). The workflow timer measures end-to-end execution, including model calls, dependency waits, client communication, and local tool processing; it excludes server startup and condition initialization.
}

\subsection{Workflow Definitions}

\revHMGh{
We evaluate five fixed workflow structures:
\begin{itemize}
    \item \textbf{Star:} Fourteen parallel workers independently process long-context inputs, followed by one aggregation call (15 model calls in total).
    \item \textbf{Chain:} Eight sequential model calls, with later stages operating on compressed intermediate context (8 calls).
    \item \textbf{DAG:} Two dependency waves, each containing six parallel worker calls, followed by a reducer (13 calls).
    \item \textbf{Tool-ReAct:} Eight agents, each performing eight sequential model steps interleaved with local JSON-processing operations (64 calls).
    \item \textbf{Debate:} Two rounds involving two debater calls per round, followed by a synthesis/judge call (5 calls).
\end{itemize}
The dependency structures and call counts are fixed across all experimental conditions. Star, DAG, and Chain use a 64-token per-call output cap; Tool-ReAct uses a 16-token cap and local processing steps; Debate uses a 256-token cap. These workflows deliberately differ in their degrees of parallelism, sequential dependencies, context requirements, and local computation.
}

\subsection{Serving Conditions}

\revHMGh{
We evaluate five serving conditions that impose different resource constraints (Table~\ref{tab:motivation_conditions}). Clean provides unconstrained single-GPU serving, while the remaining conditions restrict KV-cache capacity, sequence admission, CPU resources, background GPU load, or combinations of these constraints.
}

\begin{table}[htbp]
\centering
{\color{revBlue}
\caption{\revHMGh{Serving configurations for the controlled workflow-ranking experiment in Fig.~1(b).}}
\label{tab:motivation_conditions}
\small
\begin{tabular}{@{}p{0.19\linewidth}p{0.75\linewidth}@{}}
\toprule
\textbf{Condition} & \textbf{Configuration} \\
\midrule
Clean & 32 GiB KV cache; \texttt{max\_num\_seqs=16}; no injected CPU or background-request pressure. \\
KV-tight & 2 GiB KV cache; \texttt{max\_num\_seqs=1}. \\
CPU-tight & Server restricted to one CPU core, with three competing CPU workers. \\
GPU pressure & Background generation requests submitted at a 20\,ms polling interval, subject to an active-request cap of 8 and a 1{,}536-token output cap. \\
Mixed-tight & 2 GiB KV cache; \texttt{max\_num\_seqs=1}; single-core CPU restriction with one competing worker; background requests at a 500\,ms polling interval, active-request cap 1, and 256-token output cap. \\
\bottomrule
\end{tabular}
}
\end{table}

\revtaDc{
The comparison holds workflow definitions, synthetic context, model configuration, and decoding settings fixed while varying serving conditions. Consequently, Fig.~1(b) provides a controlled motivation for examining workflow--substrate interactions, whereas the main benchmark extends this investigation to heterogeneous MAS tasks and PlanAgent-generated candidates across serving cells and offered loads.
}

\subsection{Repeated Executions and Latency Uncertainty}

\begin{figure*}[htbp]
    \centering
    \includegraphics[width=\textwidth]{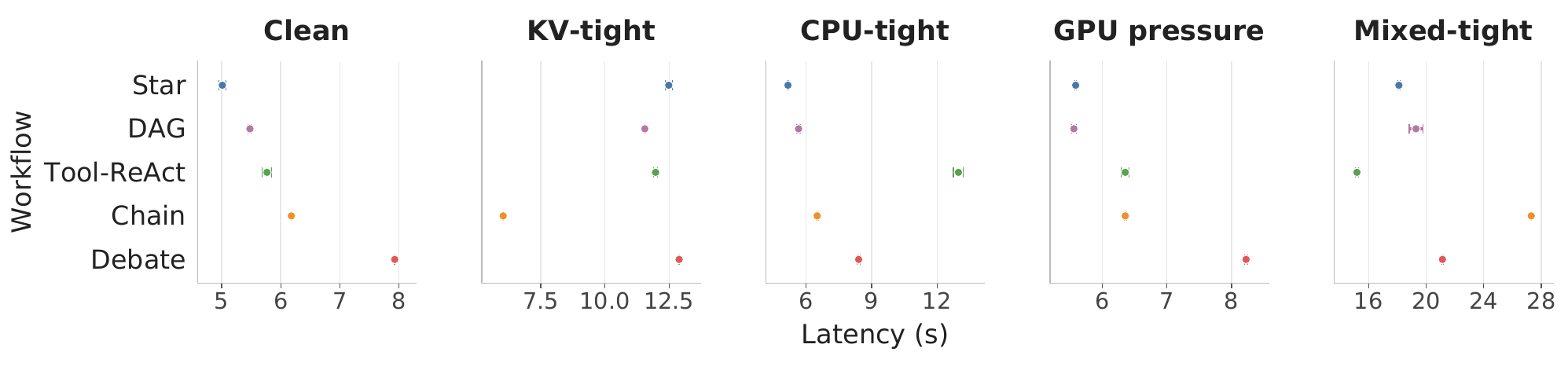}
    \caption{\revxXpx{\textbf{End-to-end latency of five fixed workflows under five serving conditions.} Points indicate mean latency, and horizontal error bars show nominal 95\% Student-$t$ confidence intervals across ten repeated executions per workflow--condition pair ($n=10$). All repetitions within a condition share one serving instance, characterizing within-instance variability under fixed workflow definitions. Each panel uses a separate latency scale for readability.}}
    \label{fig:workflow_latency_ci}
\end{figure*}

\revxXpx{
The original Fig.~1(b) experiment executed each of the 25 workflow--condition combinations ten times, producing 250 workflow executions. For each serving condition, one serving instance was launched and reused across all ten repetition blocks. Each block executed the five workflows sequentially in a deterministically shuffled order matched across conditions for the same repetition ID. This closed-loop diagnostic contains no Poisson arrival process or arrival resampling. Figure~\ref{fig:workflow_latency_ci} summarizes mean end-to-end latency and nominal within-instance 95\% Student-$t$ confidence intervals across all 25 workflow--condition pairs.
}

\subsection{Ranking Stability Across Serving Conditions}

\revxXpx{We examine whether the workflow ranking reversals persist across repetition blocks, defining $\Delta$ as Star latency minus Chain latency. Star is faster than Chain in all ten Clean repetitions ($5.0118 \pm 0.0804$\,s versus $6.1803 \pm 0.0081$\,s; mean difference $\Delta = -1.1685$\,s, paired 95\% CI $[-1.2236,-1.1135]$\,s), whereas Chain is faster than Star in all ten KV-tight repetitions ($12.4900 \pm 0.2018$\,s for Star versus $6.0392 \pm 0.0068$\,s for Chain; $\Delta = +6.4507$\,s, CI $[6.3070,6.5945]$\,s), with completely non-overlapping latency ranges under both conditions. Table~\ref{tab:motivation_rank_stability} summarizes the five-workflow rankings across all ten repetition blocks.}

\begin{table}[htbp]
\centering
{\color{revViolet}
\caption{\revxXpx{Ranking consistency across ten repetition blocks in Fig.~1(b). Rankings are ordered from fastest to slowest; per-block Kendall's $\tau$ compares each block against the corresponding Clean block.}}
\label{tab:motivation_rank_stability}
\small
\begin{tabular}{@{}lccc@{}}
\toprule
\textbf{Condition} & \textbf{Aggregate rank} & \textbf{Matches} & \textbf{Per-block $\boldsymbol{\tau}$ range} \\
\midrule
Clean & S/D/T/C/B & 10/10 & 1.00 \\
KV-tight & C/D/T/S/B & 10/10 & 0.00 \\
CPU-tight & S/D/C/B/T & 10/10 & 0.60 \\
GPU pressure & D/S/T/C/B & 5/10 & 0.60--1.00 \\
Mixed-tight & T/S/D/B/C & 9/10 & 0.20--0.40 \\
\bottomrule
\end{tabular}

\vspace{2pt}
{\footnotesize S: Star; D: DAG; T: Tool-ReAct; C: Chain; B: Debate.}
}
\end{table}

\revxXpx{
Although closely spaced middle-rank workflows exhibit minor swaps under GPU and mixed pressure, the Clean/KV-tight Star--Chain reversal holds in 10 of 10 blocks, confirming repeatability under controlled workflow execution without arrival resampling (measured within one serving instance per condition without server restarts between repetitions).
}

\section{\texorpdfstring{\revHMGh{Multi-Seed Repetitions and Statistical Uncertainty}}{Multi-Seed Repetitions and Statistical Uncertainty}}
\label{app:multiseed}

\subsection{Experimental Protocol, Realized Counts, and Statistical Method}

\revHMGh{
To evaluate variability across independent workload realizations beyond the original seed 42 sweeps, we executed seeds 43 and 44 across all five serving cells and five offered loads (0.005, 0.01, 0.02, 0.05, and 0.10 jobs/s) on 2$\times$H200 (50 additional runs; 75 runs total on 2$\times$H200), and across the three native serving cells (vLLM, SGLang, and Continuum) at 0.05 and 0.10 jobs/s on 4$\times$H200 (12 additional runs; 18 runs total for those cells). Each seed independently resamples the task mixture and Poisson arrival times, while all serving cells within a given hardware--load--seed block replay identical candidate plans and scheduled arrival offsets (Table~\ref{tab:realized_job_counts}). All runs enforce the same 30-minute submission window and 1{,}200\,s per-job deadline.
}

\begin{table}[htbp]
\centering
{\color{revBlue}
\caption{\revHMGh{Realized per-cell Poisson job counts across workload seeds 42, 43, and 44 (reported as \texttt{s42/s43/s44}). Within each hardware--load--seed block, every serving cell replays the exact same job sequence. On 4$\times$H200, sweeps start at 0.01 JPS.}}
\label{tab:realized_job_counts}
\small
\setlength{\tabcolsep}{3pt}
\resizebox{\linewidth}{!}{
\begin{tabular}{@{}llcccccc@{}}
\toprule
\textbf{JPS} & \textbf{Regime} & \textbf{Total} & \textbf{SWE-Bench} & \textbf{ResearchBench} & \textbf{DeepFund} & \textbf{PinchBench} & \textbf{PaperBanana} \\
\midrule
0.005 & 2$\times$H200 & 11 / 7 / 14 & 2 / 2 / 5 & 5 / 0 / 2 & 1 / 3 / 1 & 1 / 1 / 5 & 2 / 1 / 1 \\
0.01  & 2$\times$/4$\times$H200 & 23 / 20 / 25 & 3 / 4 / 8 & 7 / 3 / 3 & 6 / 7 / 7 & 4 / 1 / 5 & 3 / 5 / 2 \\
0.02  & 2$\times$/4$\times$H200 & 42 / 40 / 51 & 6 / 8 / 11 & 13 / 8 / 7 & 9 / 13 / 16 & 7 / 3 / 11 & 7 / 8 / 6 \\
0.05  & 2$\times$/4$\times$H200 & 85 / 101 / 107 & 15 / 27 / 23 & 22 / 21 / 20 & 22 / 18 / 24 & 13 / 18 / 20 & 13 / 17 / 20 \\
0.10  & 2$\times$/4$\times$H200 & 170 / 203 / 204 & 29 / 51 / 40 & 40 / 35 / 47 & 38 / 39 / 41 & 33 / 39 / 32 & 30 / 39 / 44 \\
\bottomrule
\end{tabular}
}
}
\end{table}

\revxXpx{
For each serving cell and load, we report the arithmetic mean and untruncated 95\% Student-$t$ confidence interval across run-level metrics, $\bar{x}\pm t_{0.975,n-1}s/\sqrt{n}$, with $n=3$ ($df=2$) unless noted otherwise. Run-level P95 end-to-end runtime is computed separately within each run over eligible non-deadline-killed execution-successful jobs before summarizing across seeds. Effective quality is calculated per family using that run's submitted-job count as the denominator. Conventional untruncated Student-$t$ endpoints that extend outside $[0,100\%]$ or $[0,1]$ reflect small run-level degrees of freedom ($df=2$) and sparse low-load family samples; at 0.005 jobs/s, ResearchBench uses $n=2$ ($df=1$) because seed 43 sampled zero ResearchBench jobs.
}

\subsection{2\texorpdfstring{$\times$}{x}H200 Completion and Tail Runtime}

\begin{figure}[htbp]
    \centering
    \includegraphics[width=1\linewidth]{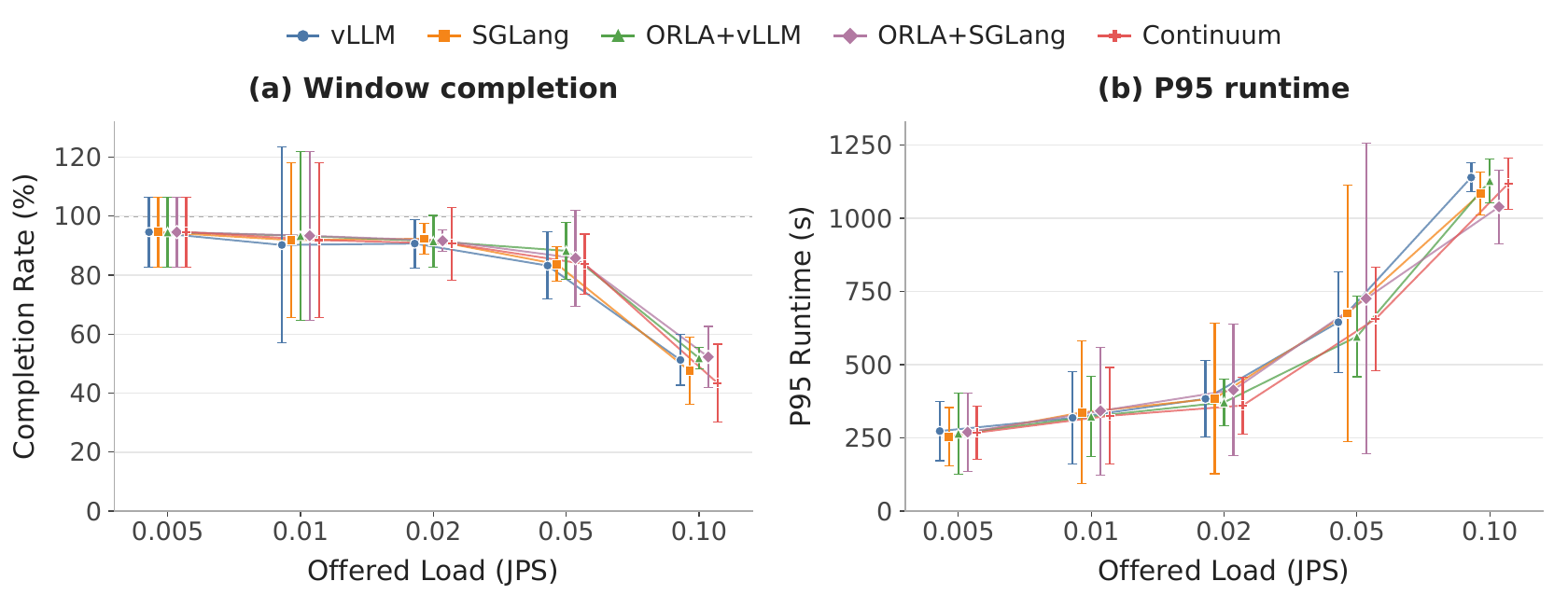}
    \caption{\revHMGh{\textbf{Multi-seed completion and tail-runtime measurements under 2$\times$H200 across five serving cells and five offered loads.} Points show the mean across three workload seeds (42--44), and error bars represent untruncated 95\% Student-$t$ confidence intervals ($n=3$). (a) Window completion rate. (b) Mean of run-level P95 end-to-end runtimes, computed within each run prior to cross-seed aggregation.}}
    \label{fig:multiseed_service}
\end{figure}

\revHMGh{Figure~\ref{fig:multiseed_service} shows that across all five serving cells on 2$\times$H200, mean run-level P95 runtime increases monotonically from 253.8--273.6\,s at 0.005 jobs/s to 1{,}039.1--1{,}139.6\,s at 0.10 jobs/s, while mean window completion drops by 42.3--51.2 percentage points.} \revtaDc{Wider intervals at 0.005--0.01 jobs/s reflect sparse Poisson realizations, whereas intervals tighten substantially at 0.05--0.10 jobs/s (85--204 jobs per cell).}

\subsection{2\texorpdfstring{$\times$}{x}H200 Per-Family Effective Quality}

\begin{figure}[!htbp]
    \centering
    \includegraphics[width=0.96\linewidth]{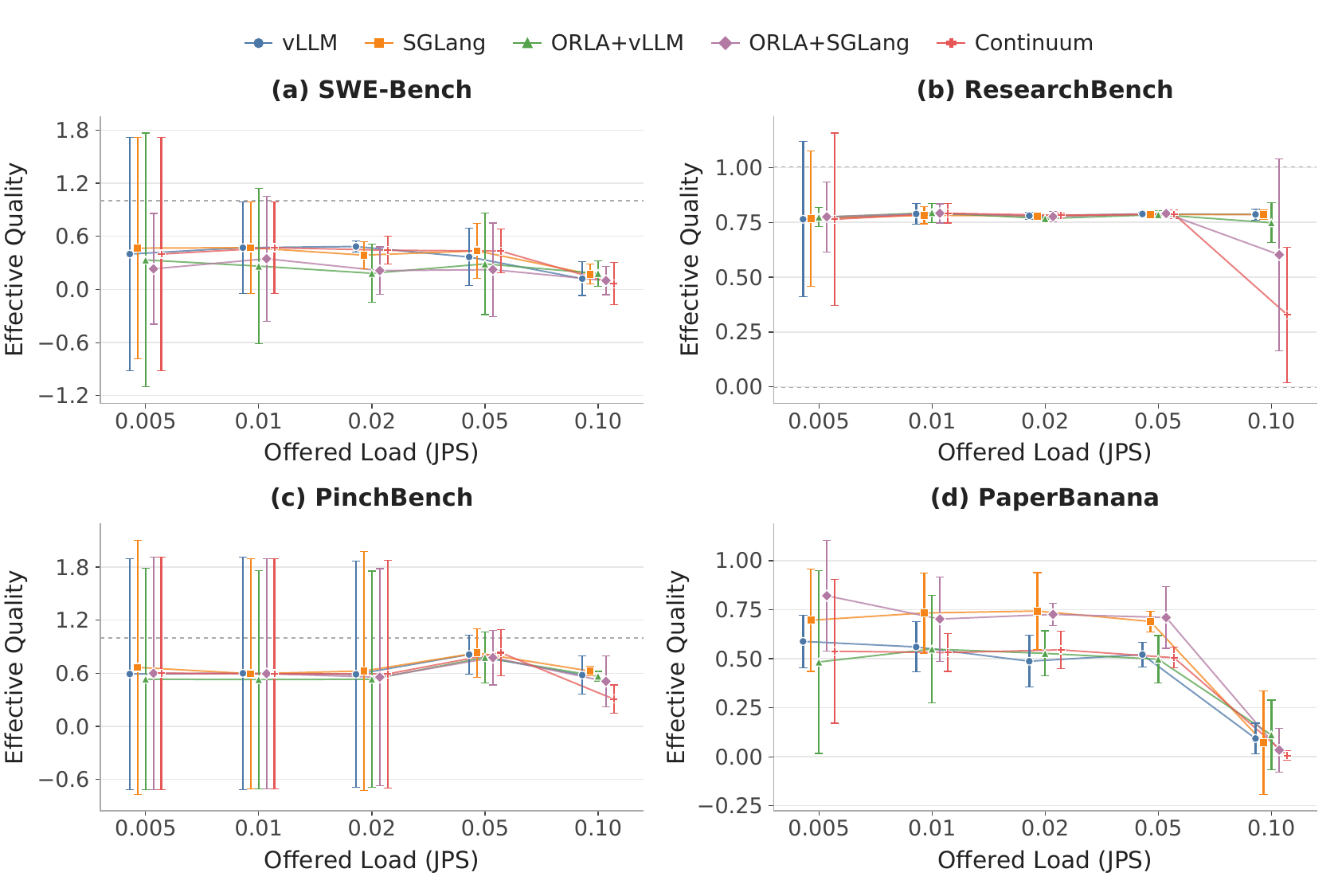}
    \caption{\revxXpx{\textbf{Per-family effective quality under 2$\times$H200 across five serving cells and five offered loads.} Points indicate the mean across three workload seeds (42--44), and error bars show untruncated 95\% Student-$t$ confidence intervals ($n=3$, except ResearchBench at 0.005 JPS where $n=2$ because seed 43 sampled zero jobs from that family).}}
    \label{fig:multiseed_quality}
\end{figure}

\revxXpx{
Figure~\ref{fig:multiseed_quality} presents per-family effective quality across seeds 42--44 on 2$\times$H200 (omitting DeepFund because its market-dependent return label is not used as a quality-robustness target). SWE-Bench, PinchBench, and PaperBanana exhibit substantial effective-quality degradation at 0.10 jobs/s across cells, whereas ResearchBench shows backend-dependent sensitivity; for example, Continuum's ResearchBench effective quality drops from 0.787 to 0.329 while its evaluated-only quality remains nearly unchanged (0.787 to 0.784), confirming that delivery loss drives the drop.
}

\subsection{Paired Changes between Moderate and High Load on 2\texorpdfstring{$\times$}{x}H200}

\revtaDc{
To quantify degradation from 0.05 to 0.10 jobs/s within each serving cell, Table~\ref{tab:multiseed_load_deltas} pairs run-level measurements by workload seed ($\Delta = \mathrm{metric}_{0.10}-\mathrm{metric}_{0.05}$, $n=3$ paired seed differences per cell).
}

\begin{table}[htbp]
\centering
{\color{revOrange}
\caption{\revtaDc{Paired changes from 0.05 to 0.10 jobs/s on 2$\times$H200 across three paired workload seeds (mean change and 95\% Student-$t$ CI). Negative completion changes denote degradation; positive P95 changes denote longer tail runtime.}}
\label{tab:multiseed_load_deltas}
\small
\begin{tabular}{@{}lcc@{}}
\toprule
\textbf{Serving cell} & \textbf{$\Delta$ Completion (pp)} & \textbf{$\Delta$ Run-level P95 (s)} \\
\midrule
vLLM        & $-31.98$ $[-38.69,-25.26]$ & $+494.4$ $[298.7,690.2]$ \\
SGLang      & $-36.21$ $[-41.86,-30.56]$ & $+408.6$ $[-71.2,888.3]$ \\
Continuum   & $-40.38$ $[-43.45,-37.31]$ & $+462.5$ $[198.7,726.3]$ \\
ORLA+vLLM   & $-36.32$ $[-47.38,-25.27]$ & $+531.8$ $[443.8,619.8]$ \\
ORLA+SGLang & $-33.48$ $[-41.23,-25.73]$ & $+313.0$ $[-208.5,834.6]$ \\
\bottomrule
\end{tabular}
}
\end{table}

\revHMGh{
Across all 15 matched serving-cell/seed comparisons on 2$\times$H200, completion falls (mean drops of 31.98--40.38 percentage points; all five paired CIs exclude zero) and run-level P95 runtime rises (+313.0 to +531.8\,s; three of five CIs exclude zero).} \revxXpx{PaperBanana and PinchBench effective quality likewise decline in all 15 comparisons and SWE-Bench declines in 13 of 15. These paired contrasts establish consistent within-cell degradation under load, while cross-cell comparisons characterize multi-dimensional service envelopes without claiming pairwise statistical separation.}

\subsection{4\texorpdfstring{$\times$}{x}H200 Multi-Seed Replication}

\revHMGh{
We further replicated the 0.05 and 0.10 jobs/s sweeps on 4$\times$H200 across seeds 42--44 for the three native backends (vLLM, SGLang, and Continuum; 18 runs totaling 2{,}610 job executions). For seeds 43 and 44, candidate plans were generated once for the four-card model pool by PlanAgent and replayed identically across all three backends within each seed--load block (verified via exact input fingerprints). Table~\ref{tab:multiseed_4h_summary} summarizes the resulting service and survivor-matched tail-latency metrics.
}

\begin{table*}[htbp]
\centering
{\color{revBlue}
\caption{\revHMGh{Three-seed service and matched-cohort latency summary on 4$\times$H200 (seeds 42--44; mean and 95\% Student-$t$ CI). ``Matched $\Delta$P95'' restricts each seed to shared-task prefix jobs (identical job IDs and tasks across 0.05 and 0.10 JPS) that succeed on all three backends ($n=68, 74, 82$ jobs in seeds 42, 43, 44), isolating queueing inflation from survivor-selection shifts.}}
\label{tab:multiseed_4h_summary}
\small
\setlength{\tabcolsep}{4pt}
\renewcommand{\arraystretch}{1.10}
\begin{tabular*}{\linewidth}{@{\extracolsep{\fill}}lclll@{}}
\toprule
\textbf{Backend} & \textbf{Load (JPS)} & \textbf{Completion (\%)} & \textbf{Execution success (\%)} & \textbf{Run-level P95 (s)} \\
\midrule
vLLM      & 0.05 & 88.55 [78.77, 98.33]  & 98.32 [97.24, 99.41]  & 412.3 [350.7, 473.9] \\
          & 0.10 & 82.63 [67.53, 97.73]  & 96.17 [89.19, 103.15] & 814.7 [456.2, 1173.3] \\
\addlinespace[2pt]
SGLang    & 0.05 & 89.82 [75.13, 104.51] & 95.00 [86.41, 103.59] & 651.1 [$-143.7$, 1445.9] \\
          & 0.10 & 82.52 [73.86, 91.18]  & 81.26 [49.70, 112.82] & 377.7 [130.9, 624.4] \\
\addlinespace[2pt]
Continuum & 0.05 & 89.21 [82.13, 96.28]  & 98.08 [93.93, 102.22] & 407.2 [301.0, 513.3] \\
          & 0.10 & 80.60 [61.52, 99.68]  & 92.51 [71.70, 113.32] & 866.5 [322.6, 1410.4] \\
\midrule
\textbf{Backend} & \textbf{Paired $\Delta$P95 (s)} & \textbf{Paired 95\% CI (s)} & \textbf{Matched $\Delta$P95 (s)} & \textbf{Seeds with increase} \\
\midrule
vLLM      & $+402.5$ & $[+104.3, +700.6]$  & $+152.0$ & 3/3 \\
SGLang    & $-273.4$ & $[-1315.0, +768.1]$ & $+244.0$ & 3/3 \\
Continuum & $+459.4$ & $[-16.9, +935.7]$   & $+302.9$ & 3/3 \\
\bottomrule
\end{tabular*}
}
\end{table*}

\revHMGh{
The 4$\times$H200 replications reinforce two central benchmark conclusions. First, expanded multi-model capacity mitigates quality loss on SWE-Bench (effective quality $0.458 \to 0.457$ on vLLM; $0.446 \to 0.498$ on Continuum) and ResearchBench ($0.806 \to 0.802$--$0.805$), while interactive PinchBench effective quality still declines from $0.797$--$0.815$ to $0.669$--$0.762$. Second, Table~\ref{tab:multiseed_4h_summary} provides a direct multi-seed demonstration of survivorship bias in unconditioned latency metrics: under SGLang at 0.10 jobs/s in seeds 43 and 44, $35/39$ and $39/44$ multimodal PaperBanana jobs hit the 1{,}200\,s deadline (lowering SGLang's mean PaperBanana effective quality from $0.766$ to $0.300$ while evaluated-only quality stays flat at $0.766 \to 0.764$). Because deadline-terminated jobs exit the survivor P95 cohort, SGLang's raw run-level P95 appears lower at 0.10 jobs/s ($377.7$\,s) than vLLM ($814.7$\,s, paired $\Delta = +402.5$\,s, CI $[+104.3,+700.6]$\,s) or Continuum ($866.5$\,s), which achieve $96.17\%$ and $92.51\%$ execution success. Restricting to the six-way matched surviving-job intersection removes this selection artifact, revealing that P95 runtime increases under load in all nine backend--seed pairs (+152.0\,s on vLLM, +244.0\,s on SGLang, and +302.9\,s on Continuum).
}

\section{\texorpdfstring{\revtaDc{Workflow Ranking Analysis in the Main Benchmark}}{Workflow Ranking Analysis in the Main Benchmark}}
\label{app:main_workflow_ranking}

\subsection{Cross-Substrate Workflow Ranking Reversals}

\revtaDc{To examine whether substrate-dependent workflow ranking reversals also occur in the main benchmark, we compare Chain and finite-state-machine (FSM) workflows on PinchBench at 4$\times$H200 and 0.10 jobs/s (seed 42). To avoid backend-specific differences in successful-job selection, latency is computed using jobs successfully executed across all five serving cells for each topology (14 Chain and 7 FSM jobs). Table~\ref{tab:main_ranking_reversal} reports the mean end-to-end runtimes and topology-specific execution-success counts.}

\revtaDc{Chain achieves lower mean latency under vLLM, SGLang, and ORLA+SGLang, whereas FSM is faster under Continuum and ORLA+vLLM. Because each topology is evaluated on a matched successful-job intersection, the observed reversal is not explained by differences in which jobs enter the latency calculation across serving cells. In our 4$\times$H200 multi-seed native runs at 0.10 jobs/s, the same cross-substrate reversal between SGLang and Continuum recurs in seed 43 across 35 three-way matched PinchBench jobs (19 Chain, 16 FSM): Chain is faster than FSM under SGLang (124.90\,s versus 226.82\,s), whereas FSM is faster than Chain under Continuum (135.38\,s versus 155.95\,s) and vLLM (136.45\,s versus 169.83\,s), while seed 44 favors FSM across all three backends.}

\begin{table}[htbp]
\centering
{\color{revOrange}
\caption{\revtaDc{Chain/FSM latency ranking reversal on PinchBench at 4$\times$H200 and 0.10 jobs/s (seed 42). Latencies are computed over jobs successfully executed across all five serving cells (Chain: $n=14$; FSM: $n=7$). Success counts use all submitted jobs of each topology.}}
\label{tab:main_ranking_reversal}
\small
\setlength{\tabcolsep}{5pt}
\begin{tabular}{@{}lrrcc@{}}
\toprule
\textbf{Serving cell} & \textbf{Chain (s)} & \textbf{FSM (s)} & \textbf{Faster} & \textbf{Success (C/F)} \\
\midrule
vLLM        & 291.12 & 375.52 & Chain & 19/23, 7/10 \\
SGLang      & 116.43 & 127.56 & Chain & 23/23, 9/10 \\
Continuum   & 480.13 & 394.45 & FSM   & 14/23, 7/10 \\
ORLA+vLLM   & 324.82 & 294.07 & FSM   & 22/23, 9/10 \\
ORLA+SGLang & 193.25 & 217.71 & Chain & 23/23, 9/10 \\
\bottomrule
\end{tabular}
}
\end{table}

\subsection{Cross-Load Comparability}

\revtaDc{
We additionally examined exact candidate-plan matches between 0.05 and 0.10 jobs/s in the five-cell sweeps. Requiring the same candidate plans to execute successfully across all five serving cells at both loads leaves two matched jobs under 2$\times$H200 and five under 4$\times$H200. We therefore use the matched PinchBench comparisons above to evaluate cross-substrate topology reversals at fixed load, and use the broader multi-seed sweeps to evaluate load sensitivity in completion, tail latency, and effective quality.
}

\section{\texorpdfstring{\revHMGh{Evaluation Audit and Quality Sensitivity Analysis}}{Evaluation Audit and Quality Sensitivity Analysis}}
\label{app:quality_audit}

\subsection{Decomposing Execution and Evaluation Failures}

\revHMGh{
To distinguish delivery failures from evaluation-infrastructure failures, we audited the execution and evaluation outcomes of all 3{,}255 submitted jobs in the original 45 benchmark runs. We distinguish valid numerical evaluations, jobs skipped after execution failure or deadline termination, and evaluation failures occurring after successful execution. A valid evaluation with a quality score of zero is counted as an evaluated job, not an evaluation failure.
}

\begin{table}[htbp]
\centering
{\color{revBlue}
\caption{\revHMGh{Execution and evaluation outcome audit across the original 45 runs. ``Skipped'' denotes jobs not evaluated following execution failure or deadline termination; ``Eval.\ fail'' denotes evaluation failure after successful execution. No additional evaluations were missing.}}
\label{tab:evaluation_audit}
\small
\setlength{\tabcolsep}{4pt}
\begin{tabular}{@{}lrrrrr@{}}
\toprule
\textbf{Task family} & \textbf{Submitted} & \textbf{Success} & \textbf{Valid} & \textbf{Skipped} & \textbf{Eval.\ fail} \\
\midrule
SWE-bench     & 540 & 326 & 324 & 214 & 2 \\
ResearchBench & 845 & 815 & 815 & 30  & 0 \\
DeepFund      & 755 & 752 & 752 & 3   & 0 \\
PinchBench    & 575 & 485 & 485 & 90  & 0 \\
PaperBanana   & 540 & 387 & 387 & 153 & 0 \\
\midrule
\textbf{Total} & \textbf{3,255} & \textbf{2,765} & \textbf{2,763} & \textbf{490} & \textbf{2} \\
\bottomrule
\end{tabular}
}
\end{table}

\revHMGh{
As shown in Table~\ref{tab:evaluation_audit}, 492 submitted jobs did not receive valid evaluations: 490 (99.59\%) followed execution failure or the enforced 1{,}200\,s deadline, whereas only two (0.41\%) encountered evaluation failures after successful execution. Thus, 2{,}763 of 2{,}765 execution-successful jobs (99.93\%) received valid numerical scores (and zero post-success evaluation errors occurred across all 12 new 4$\times$H200 multi-seed runs). Execution and delivery failures therefore account for almost the entire missing-score component of the effective-quality gap.
}

\revHMGh{
The two post-success evaluation failures occurred on SWE-bench instances \texttt{matplotlib\_\_matplotlib-25433} and \texttt{matplotlib\_\_matplotlib-23913} in the 4$\times$H200/ORLA+vLLM/0.10 jobs/s cell, where uncached Docker images could not be retrieved due to network errors. Excluding these two submissions shifts that cell's SWE-bench effective quality only slightly from $12/29=0.4138$ to $12/27=0.4444$, without changing any backend ranking in Fig.~\ref{fig:exp-task-quality}.
}

\subsection{Sensitivity to Execution-Grounded Quality Metrics}

\revxXpx{
We further examine whether the distinction between evaluated-only and effective quality persists without LLM-based quality judging. We consider SWE-bench, evaluated using executable Docker test suites, and the automated-only subset of PinchBench, whose graders inspect generated files and task artifacts without an LLM judge (excluding hybrid and judge-only tasks, as well as DeepFund's market-dependent labels). For $N$ submitted jobs and $V$ valid evaluations with scores $q_j$, effective quality is $Q_{\mathrm{eff}}=\sum_{j\in\mathcal{V}}q_j/N$, whereas evaluated-only quality is $Q_{\mathrm{eval}}=\sum_{j\in\mathcal{V}}q_j/V$ for $V>0$. Table~\ref{tab:grounded_quality} reports both metrics pooled across the five serving cells within each 2$\times$H200/load block of the original benchmark.
}

\begin{table}[htbp]
\centering
{\color{revViolet}
\caption{\revxXpx{Sensitivity analysis restricted to execution- or artifact-grounded quality metrics, pooled across the five serving cells within each original 2$\times$H200/load block.}}
\label{tab:grounded_quality}
\small
\setlength{\tabcolsep}{4pt}
\begin{tabular}{@{}lcccc@{}}
\toprule
\textbf{Task subset} & \textbf{JPS} & \textbf{Eval./Sub.} & $\boldsymbol{Q_{\mathrm{eff}}}$ & $\boldsymbol{Q_{\mathrm{eval}}}$ \\
\midrule
SWE-bench          & 0.05 & 57/75  & 49.33\% & 64.91\% \\
SWE-bench          & 0.10 & 28/145 & 11.72\% & 60.71\% \\
PinchBench (auto.) & 0.05 & 35/35  & 93.81\% & 93.81\% \\
PinchBench (auto.) & 0.10 & 48/65  & 66.15\% & 89.58\% \\
\bottomrule
\end{tabular}
}
\end{table}

\revxXpx{
From 0.05 to 0.10 jobs/s, effective quality decreases by 37.61 percentage points for SWE-bench and 27.66 points for automated-only PinchBench, whereas evaluated-only quality decreases by only 4.20 and 4.23 points, respectively. At 0.10 jobs/s, the 117 unevaluated SWE-bench submissions comprise 103 deadline terminations and 14 generation/execution failures, and all 17 unevaluated automated-only PinchBench submissions were deadline-terminated, with zero post-success evaluation errors. This gap varies by condition and is not universal: for example, 4$\times$H200/Continuum/0.10 jobs/s evaluates all 29 SWE-bench submissions ($Q_{\mathrm{eff}} = Q_{\mathrm{eval}} = 17/29 = 58.62\%$). Overall, restricting evaluation to judge-free grounded metrics preserves our central finding that evaluated-only scoring can substantially overstate useful quality delivered per submission.
}